\documentclass[a4paper,11pt]{article}
\pdfoutput=1 

\usepackage{jinstpub} 
\usepackage{comment}
\usepackage{float}
\usepackage{url}
\title{Material Identification in Nuclear Waste Drums using Muon Scattering Tomography and Multivariate Analysis}
 
\author[a,1]{M.J.~Weekes,\note{Corresponding author.}}
\author[a]{A.F.~Alrheli,}
\author[a]{D.~Barker,}
\author[b]{D.~Kiko\l{}a,}
\author[c]{A.K.~Kopp,}
\author[b,c]{M.~Mhaidra,}
\author[a]{J.P.~Stowell,}
\author[a]{L.F.~Thompson,}
\author[c]{J.J.~Velthuis.}

\affiliation[a]{University of Sheffield, Department of Physics and Astronomy, Hounsfield Road, Sheffield, S3 7RH, UK.}
\affiliation[b]{Warsaw University of Technology, Pl. Politechniki 1, 00-661 Warsaw, Poland}
\affiliation[c]{University of Bristol, School of Physics, HH Wills Physics Laboratory,
Tyndall Avenue, Bristol BS8 1TL, UK.}

\emailAdd{mweekes1@sheffield.ac.uk}

\abstract{
The use of muon scattering tomography for the non-invasive characterisation of nuclear waste is well established. We report here on the application of a combination of feature discriminators and multivariate analysis techniques to locate and identify materials in nuclear waste drums. After successful training and optimisation of the algorithms they are then tested on a range of material configurations to assess the system's performance and limitations. The system is able to correctly identify uranium, iron and lead objects on a ~few \text{cm} scale. The system's sensitivity to small uranium objects is also established as $0.90^{+0.07}_{-0.12}$, with a false positive rate of $0.12^{+0.12}_{-0.07}$.}

\keywords{Analysis and Statistical Methods, Particle Tracking Detectors}

\begin{document}

\maketitle
\flushbottom
\newpage
\section{Introduction}
\label{sec:intro}
It is important to develop non-destructive methods to determine the contents of sealed nuclear waste packages, in order to minimise the risks of environmental contamination and personnel radiation exposure and to allow for more effective safeguarding. Non-Destructive Assay (NDA) techniques in current use include calorimetry and Muon Scattering Tomography (MST).\par
NDA techniques can analyse drum contents in a variety of ways. For example, calorimetry can be used to measure the mass of nuclear material inside a container by its heat emission \cite{kubinski2020calorimetric}. In contrast, MST (with exposure times of several days to weeks) can produce full 3D images of a volume of interest, allowing individual objects inside the drum to be viewed as well as giving information on their atomic number $Z$ and density \cite{schultz2004image}.\par
Simulation studies are useful tools to assess MST techniques and algorithms; the technique described in this paper was developed and tested via Monte Carlo simulations. It uses MST data in combination with Multi-Variate Analysis (MVA) classifiers and clustering algorithms to approximately identify the locations and shapes of objects stored in a concrete-filled waste drum. Subsequently, additional trained classifiers are applied to each identified object to classify them as `iron', `lead', or `uranium', representing low-threat medium-$Z$ material, low-threat high-$Z$ material, and high-threat high-$Z$ material respectively. The use of these four materials allows three classification problems of interest to be investigated: separation of stored objects from the concrete background, separating medium- and high-$Z$ materials, and distinguishing between two high-$Z$ materials.\par  Previous applications of machine learning techniques to MST imaging have demonstrated methods for distinguishing between drums containing uranium and lead blocks \cite{frazao2016discrimination} and for reconstructing the size of uranium blocks \cite{frazao2019high}. Our system builds on these through the ability to isolate and identify multiple distinct bodies of different materials and sizes in a waste drum. Other previous research into combining machine learning and MST include applications in cargo scanning \cite{stocki2012machine}\cite{pan2019experimental}, a related problem for which short exposure times are required.\par

\section{Muon scattering tomography}
Cosmic rays interact with the Earth's atmosphere to produce showers of particles, some of which subsequently decay to muons, resulting in a muon flux at sea level of around $1\ \text{cm}^{-2}\ \text{min}^{-1}$ \cite{schultz2007statistical}. These cosmic ray muons are highly penetrating due to their large mass and lack of strong interactions. They have an angular distribution that varies approximately as $\cos^2{\theta}$, where $\theta$ is the zenith angle. Muons are also highly sensitive to the atomic number $Z$ of the material they are passing through, making them suitable candidates for tomographic imaging of nuclear waste drums.\par
Muons undergo multiple elastic Coulomb scatterings in matter, with the projected scattering angles following an approximately Gaussian distribution with width $\sigma$ given by 
\begin{equation} \label{scattering angle dist width}
\sigma\approx\frac{13.6\ \text{MeV}}{\beta c p} \sqrt{X/X_0}
\end{equation}
where $\beta$ is the muon speed divided by the speed of light in a vacuum, $c$; $p$ is the muon momentum, $X$ is the thickness of the material and $X_0$ is the radiation length of the material \cite{lynch1991approximations}. The latter is given by
\begin{equation} \label{radlength}
X_0=\frac{716.4 A}{Z(Z+1)\ln (287/\sqrt{Z})}\ [\text{g}\cdot\text{cm}^{-2}]
\end{equation}
where $\rho$ is the material density and $A$ is atomic mass \cite{PDBook}.\par
\begin{figure}[htb]
\begin{center}
      \includegraphics[scale=0.6]{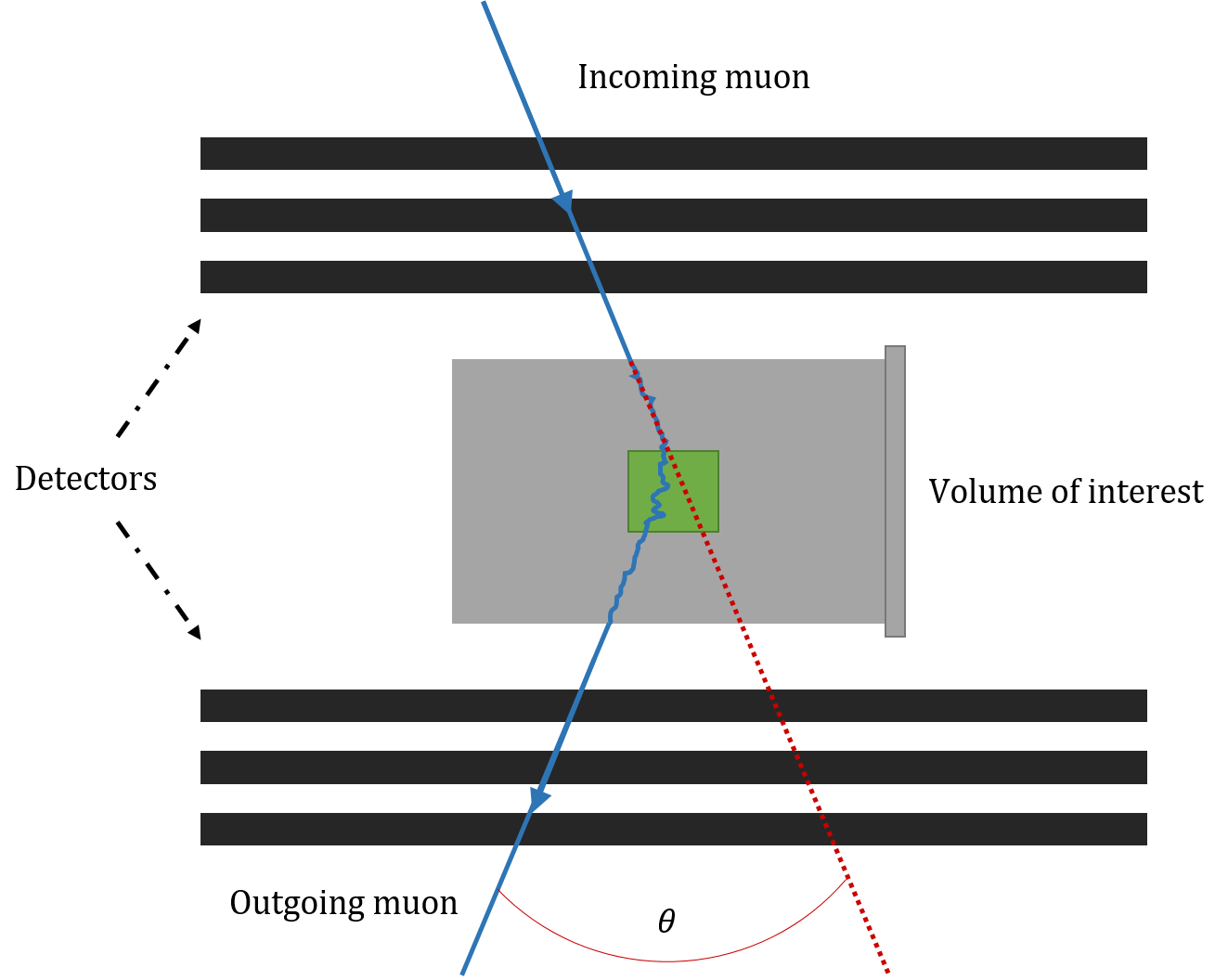}
     \caption{Schematic showing the principle of muon scattering tomography applied to a nuclear waste drum containing a block of high-$Z$ material (in green). Particle detectors measure the trajectories of muons before and after encountering the volume of interest, allowing the scattering angle $\theta$ (here exaggerated) to be calculated.} 
     \label{mst_schematic_figure}
        \end{center}    
\end{figure}

A general MST experiment consists of two sets of particle detectors, one above and one below some volume of interest such as a waste drum (see Figure \ref{mst_schematic_figure}). Multiple layers of detector are necessary in order to construct a three dimensional trajectory for each muon from the detector hits. This allows the incoming and outgoing trajectories of each muon to be measured and hence the muon scattering angles to be calculated.\par 
Several algorithms have been developed to enable imaging of a volume of interest from MST data. The simplest is the Point of Closest Approach (PoCA) algorithm \cite{riggi2013muon}, which models a muon's multiple scatterings as a single scattering at a single point (`scattering vertex'), found by extrapolating the incoming and outgoing tracks into the volume and finding the point which minimises the distance to each. This assumption allows for fast computation at the expense of image quality. A more advanced MST algorithm has been used in this study (see section \ref{BC algorithm section}) which builds on PoCA by exploiting the spatial density of scattering vertices; a high density of scattering vertices corresponds to the presence of high-$Z$ material as large-angle muon scatterings occur more often in such materials.
\subsection{Binned clustering algorithm}\label{BC algorithm section}
This algorithm, developed in \cite{thomay2013binned}, improves on the widely-used Point of Closest Approach (PoCA) muon tomography algorithm \cite{schultz2004image} by taking into account the degree of spatial clustering of muon scattering vertices. A higher density of vertices corresponds to higher-$Z$ materials (once the muon momentum is accounted for, see below) as strong muon scatterings take place with greater frequency in such materials.\par
The volume is divided into cubic voxels of side length $1\ \text{cm}$. The incoming and outgoing muon tracks are extrapolated through the volume, and the point at which the distance between the tracks is minimal (the PoCA) is designated as the scattering vertex for the muon. This is repeated for all of the detected muons that encounter the volume of interest. Next, the scattering vertices inside each $1\ \text{cm}^3$ voxel are sorted by the scattering angle of the corresponding muon, and the vertices corresponding to the $n$ largest scattering angles are kept (voxels which contain less than $n$ vertices are discarded). This factor of $n$ is an important tunable parameter of the algorithm. High values of $n$ improve the contrast between high and low-$Z$ materials, as a greater sample of muons are kept, but reduce image `quality' (i.e. the number of non-empty voxels in the image) as more voxels fall below the cut and are removed from the image.\par
\begin{figure}[htb]
\begin{center}
      \includegraphics[scale=0.3]{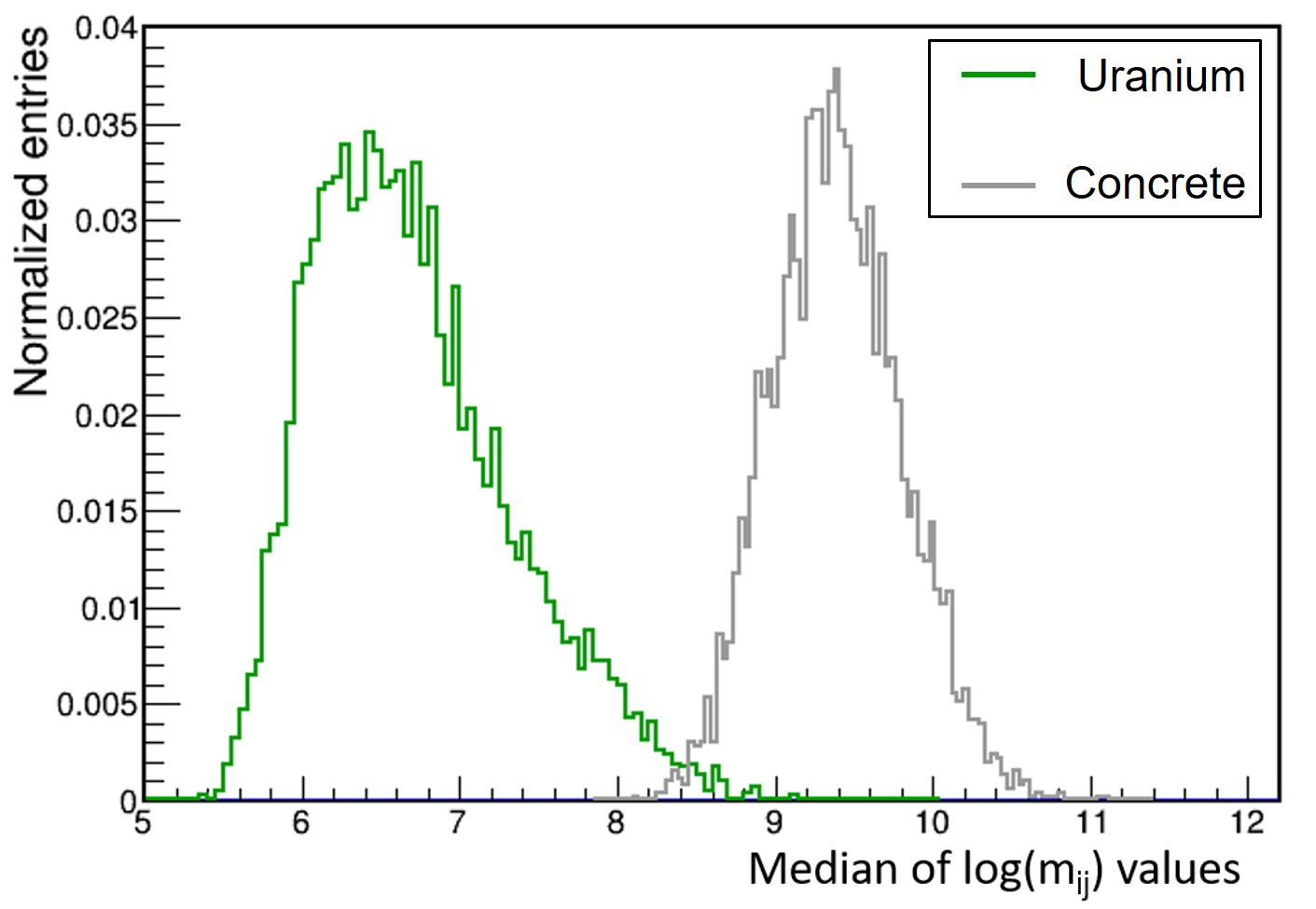}
     \caption{Comparison of distributions of the binned clustering algorithm discriminator, for $20\ \text{cm}$ cubes of uranium and concrete. Lower discriminator values correspond to higher $Z$ material.} 
      \label{ucon_dist_figure}
  \end{center}    
\end{figure}
For each of the $\binom{n}{2}$ pairs of vertices $i,j$ in each voxel, a metric value $m_{ij}$ is calculated according to
\begin{equation} \label{weighted metric}
m_{ij}=\frac{|\textbf{V}_i-\textbf{V}_j|}{(\theta_i\Tilde{p_i})\cdot(\theta_j\Tilde{p_j})}
\end{equation}
where $\textbf{V}_i$, $\theta_i$ and $\Tilde{p_i}$ are respectively the position, scattering angle and normalised (by a factor of $3\ \text{GeV/c}$) momentum of muon $i$. Weighting by muon momentum is necessary as large scattering angles could indicate low-momentum muons being scattered in low-$Z$ materials instead of strong scattering in high-$Z$ materials. In an experimental system, the muon momentum can be estimated using the muon scatterings between the detector planes, as the planes are of known material and thickness. Following the method of \cite{morris2008tomographic}, for our simulations the momentum was obtained by adding a smearing factor to the Monte Carlo truth momentum. The smearing factor was drawn from a Gaussian with width 50\% of the Monte Carlo truth momentum.\par
Finally, the median of the distribution of $\log(m_{ij})$ in a voxel is determined; this is the algorithm's discriminator value for that voxel. Comparing the distributions of this discriminator for high- and low-$Z$ materials shows that the discriminator is sensitive to $Z$ (see Figure \ref{ucon_dist_figure}).\par
\begin{figure}[htb]
\begin{center}
      \includegraphics[scale=0.55]{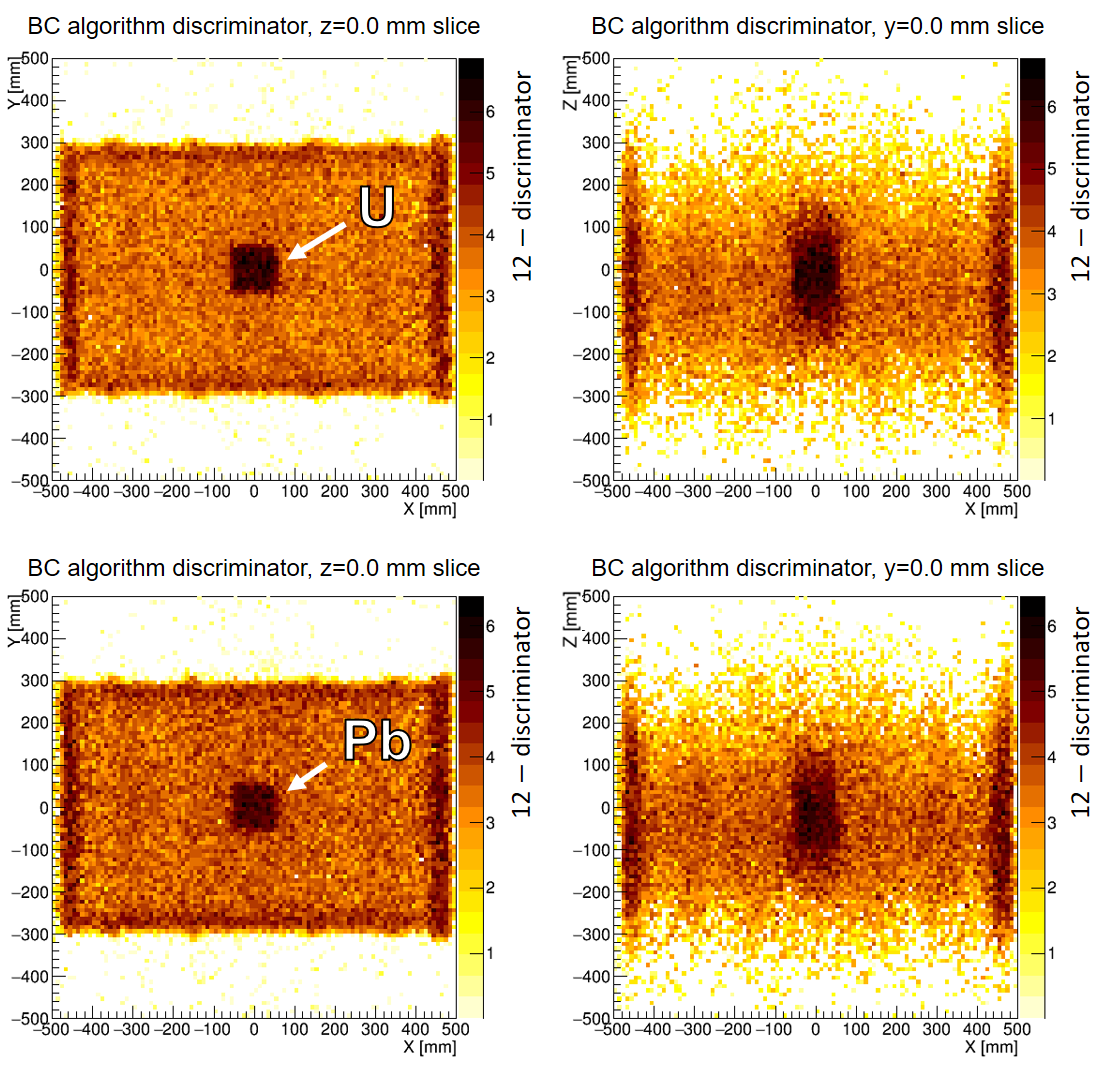}
     \caption{$xy$ (left) and $xz$ (right) slices from binned clustering output images of waste drums containing $10\ \text{cm}$ side length cubes of uranium (top) and lead (bottom). Exposure time = 10 days, $n$ = 5. The smearing effect along the $z$ axis is due to uncertainty in the scattering vertex $z$ coordinate for tracks with small scattering angles. Note that the plotted discriminator values have been subtracted from $12$ for visual clarity.} 
      \label{bc_slice_figure}
  \end{center}    
\end{figure}
For imaging purposes, each voxel is filled with its discriminator value as described above, creating a tomogram of the volume of interest. Viewing slices of discriminator values through the image (see Figure \ref{bc_slice_figure}) allows regions of high-$Z$ material to be identified visually. This gives a degree of information about the locations and morphologies of objects stored in the drum. However, it is vulnerable to a vertical smearing effect inherent in the PoCA reconstruction, and without an object of known material for comparison, it is difficult to determine the specific materials of objects `by eye'. Additionally, without any way to automatically remove background materials such as the steel drum and concrete matrix, the 3D image must be viewed in slices to determine the locations of stored objects.\par 
By default, the binned clustering algorithm only takes into account the median of the $\log(m_{ij})$ distribution in each voxel. To test the possibility that additional material information is encoded in the shape of the $\log(m_{ij})$ value distribution, variables capturing the shape were used to train MVA classifiers. These classifiers are then used to separate the regions of the image corresponding to objects stored in the drum from the concrete matrix. Subsequently the classifiers are used to assign a material to each identified object.\par 

\subsection{System configuration}
\begin{figure}[htb]
\begin{center}
      \includegraphics[scale=0.7]{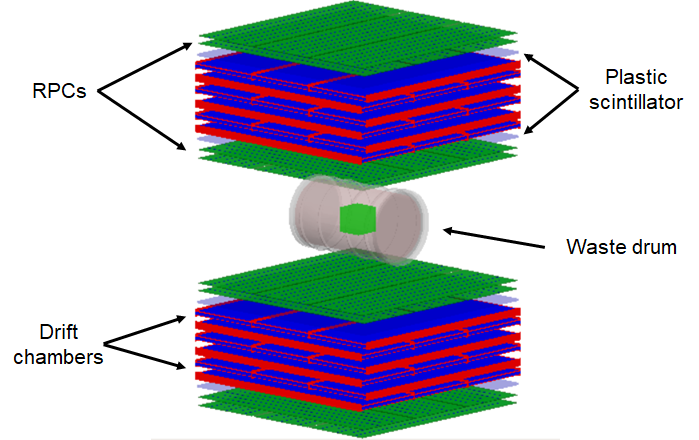}
     \caption{MST detector system simulated in CRESTA: detector modules above and below a waste drum, in which objects can be placed. The detector modules are approximately $2\ \times\ 2\ \text{m}$.} 
     \label{cresta_ucube_figure}
\end{center}    
\end{figure}

All simulations were performed using CRESTA \cite{stowellcresta}, a cosmic ray simulation platform built on the Geant4 \cite{collaboration2003geant4} particle physics toolkit and the CRY \cite{hagmann2007cosmic} cosmic ray library. Within CRESTA a MST detector system comprising two particle detector modules above and below a waste drum was simulated (see Figure \ref{cresta_ucube_figure}). This represents a `generic' MST detector system, designed for imaging a ~$1\ \text{m}$ waste drum. The detector modules are $2\ \text{m}$ by $2\ \text{m}$ and each consists of two layers of resistive plate chambers (RPCs), polystyrene scintillator triggers and three layers of drift chambers. The RPCs and drift chambers have spatial resolutions of $\sim350\ \mu\text{m}$ and $\sim2\ \text{mm}$ respectively. The detectors are arranged in alternating $x$ and $y$ layers, allowing 3D muon hits to be recorded and the incoming and outgoing tracks reconstructed.\par
\begin{figure}[htb]
\begin{center}
      \includegraphics[scale=0.4]{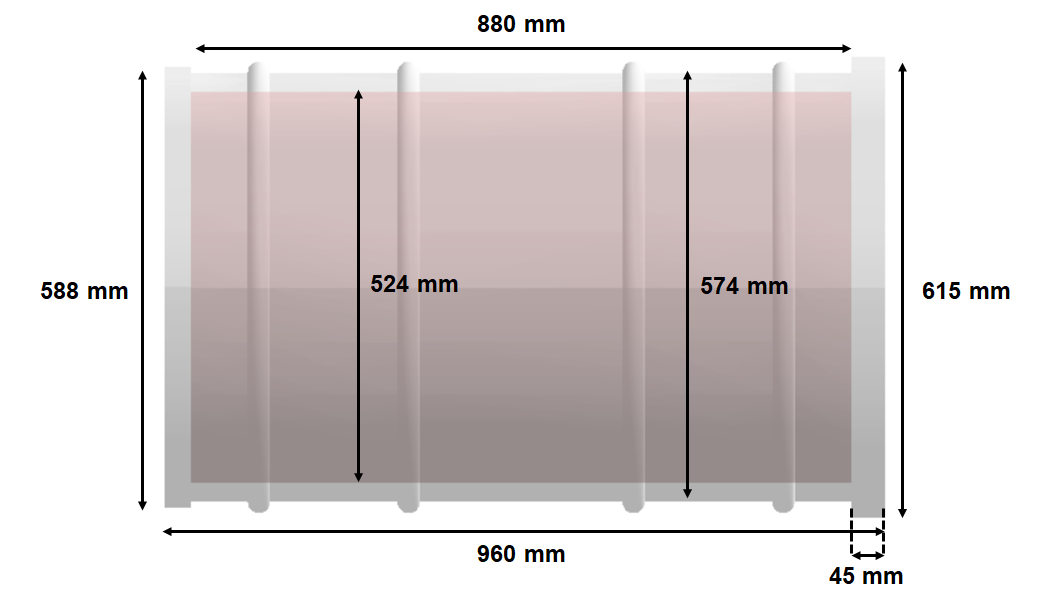}
     \caption{The simulated concrete-filled steel nuclear waste drum used in CRESTA, with its dimensions.} 
     \label{cresta_drum_figure}
        \end{center}    
\end{figure}

The waste drum is made of steel (approx. 91\% iron, 9\% carbon; element isotopes in natural abundances). It is approximately $100\ \text{cm}$ in length and $30\ \text{cm}$ in radius (see Figure \ref{cresta_drum_figure} for precise dimensions), and is filled with homogeneous concrete of density $2.3\ \text{g}\ \text{cm}^{-3}$.

\section{Multivariate analysis} \label{tmva section}
\subsection{MVAs and muon tomography}
Frazão et al. \cite{frazao2016discrimination} used MVA classifiers trained on simulated MST data to discriminate between waste drums containing lead and uranium blocks. This method can be thought of in a `global' sense, distinguishing between two categories of waste drum but not analysing the specific drum contents in terms of bodies encased in the concrete. Our approach by contrast is `local', as we are able to produce localised material information down to the scale of single $1\ \text{cm}$ voxels. This approach requires longer exposure times (of the order of several days rather than hours) but gives more detailed material information. This allows for the possibility of combining these two techniques. A user could use the former method and a short exposure to identify drums likely to contain high-$Z$ material, then subsequently apply our method and a longer exposure to the flagged drums to identify the stored objects and their materials.\par
\begin{figure}[htb]  
\begin{center}
      \includegraphics[scale=0.48]{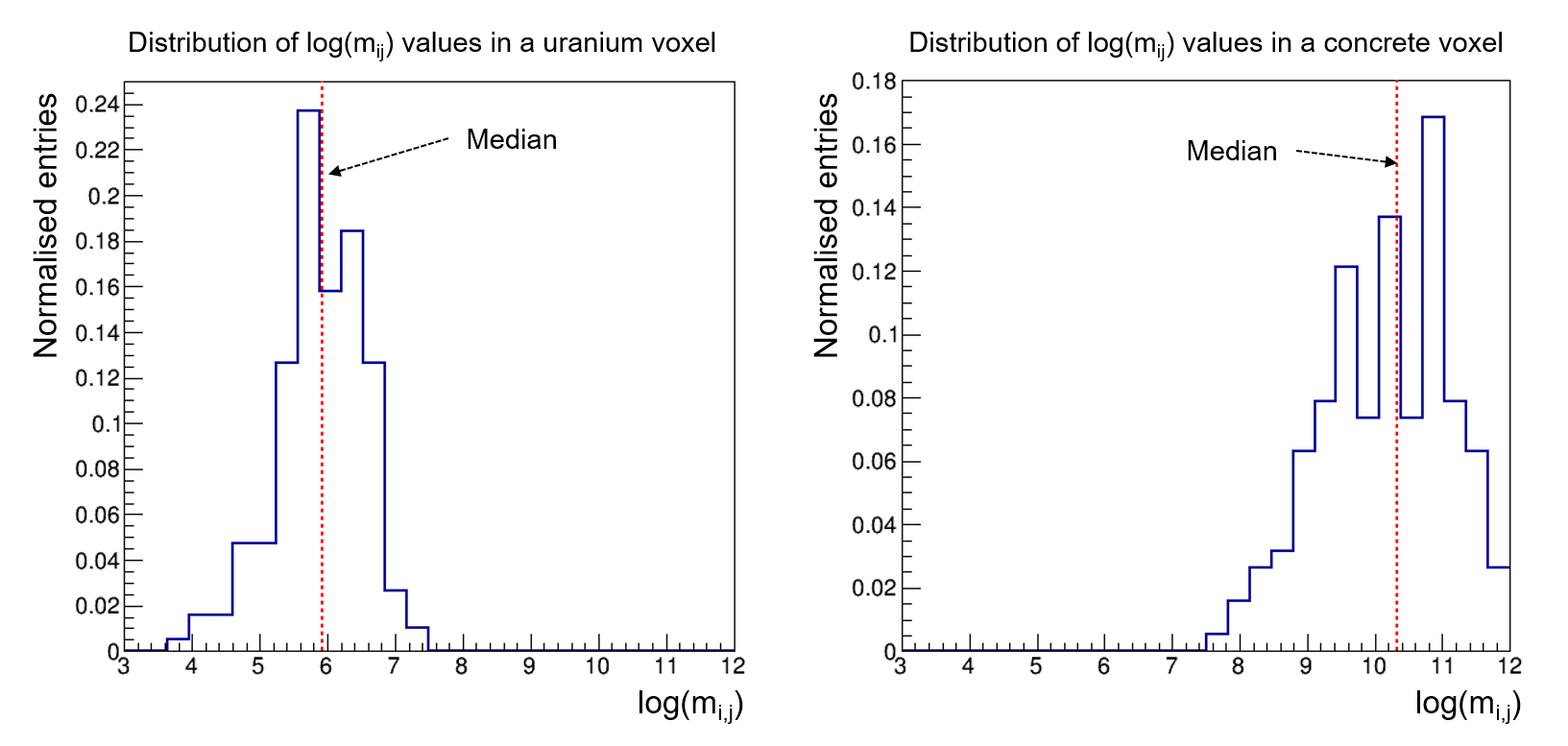}
     \caption{Comparison of distributions of $\log$(metric) values for a voxel corresponding to uranium (left) and concrete (right). The median of each distribution is used as the discriminator in binned clustering algorithm images such as Figure \ref{bc_slice_figure}. The normalised bin counts are used as the MVA input variables.} 
     \label{voxel compare figure}
  \end{center}    
\end{figure}
\begin{figure}[htb]  
\begin{center}
      \includegraphics[scale=0.39]{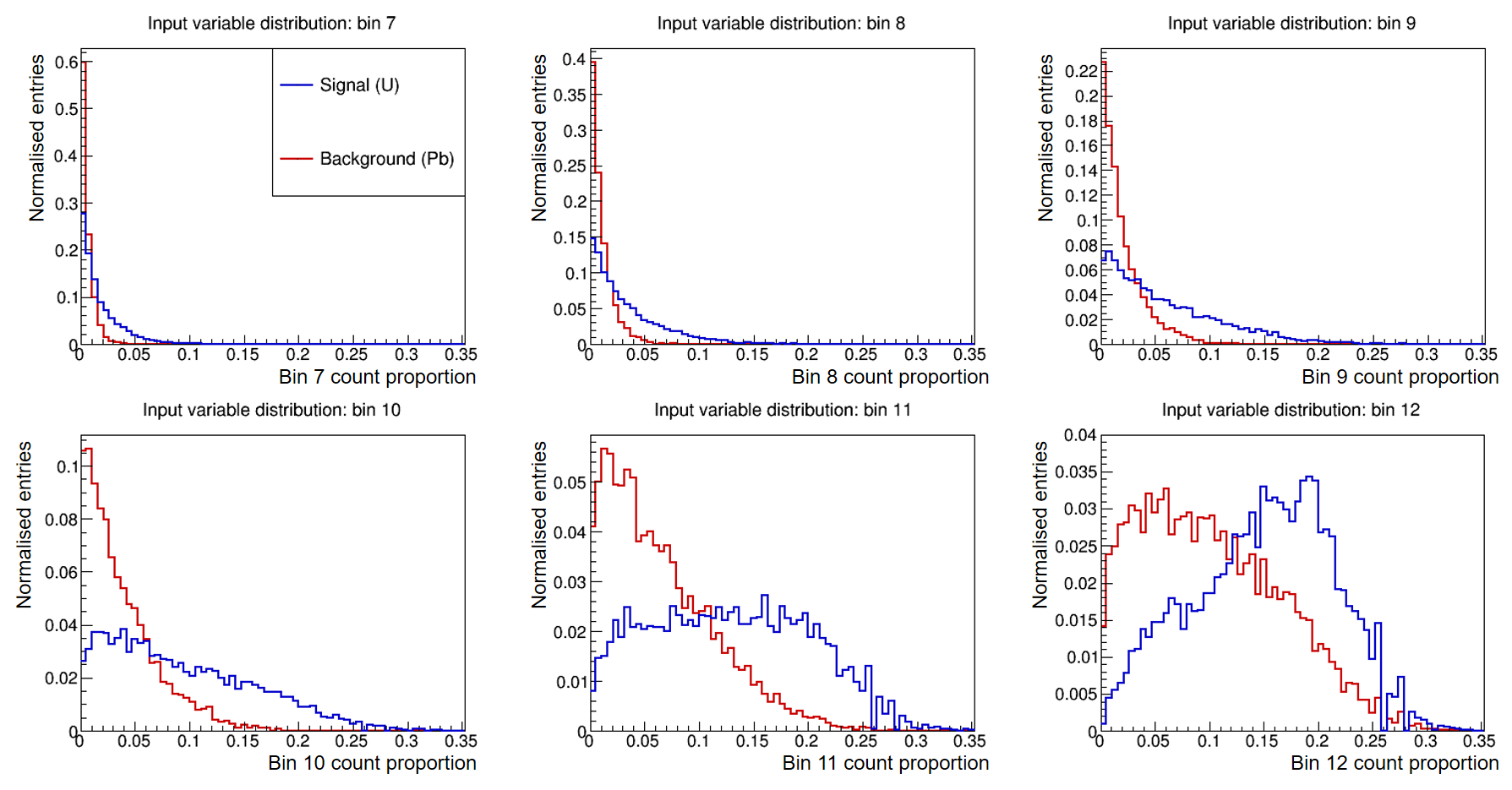}
     \caption{Example distributions of some of the input variables used to train the MVA classifiers, here specifically a binary uranium-lead classifier. The variables are the normalised bin counts (see Figure \ref{voxel compare figure}) of the $\log(m_{ij})$ values calculated by the binned clustering MST algorithm. The signal set (blue) are voxels in a $20\ \text{cm}$ cube of uranium, and the background set (red) an equivalent cube of lead.} 
     \label{tmva variables figure}
  \end{center}    
\end{figure}

Our MVA classifiers were built, trained and analysed using TMVA \cite{Hocker:2007ht}, a ROOT \cite{brun1997root}-integrated machine learning platform. Our set of variables used as input to the MVA classifiers are obtained via the binned clustering algorithm (see section \ref{BC algorithm section}). The algorithm calculates a set of metric values for each voxel, with each value corresponding to a pair of muon scattering vertices. By default, the algorithm uses the median of the distribution of $\log(m_{ij})$ values only as the discriminator for each voxel. We build on this by first binning the $\log(m_{ij})$ values into 28 bins (see Figure \ref{voxel compare figure}), calculating the normalised bin counts, and passing the counts to the MVA classifiers as the input variables (Figure \ref{tmva variables figure}) for that voxel. This approach allows more of the shape of the distribution of metric values to be captured, enhancing the information available to the classifiers. 

TMVA allows multiple MVA methods to be trained simultaneously and their efficacy compared. The performance of a binary MVA classifier can be quantified through a Receiver Operating Characteristic (ROC) curve: a plot of the true positive rate (also called sensitivity) against the false positive rate for different cuts on the classifier response for the test sample. The Area Under the Curve (AUC) is a standard measure of the classifier's discriminating power. $\text{AUC} = 1$ indicates a perfect classifier whereas an $\text{AUC} = 0.5$ would indicate the classifier performs no better than random classification.\par
\begin{figure}[htb]  
\begin{center}
      \includegraphics[scale=0.5]{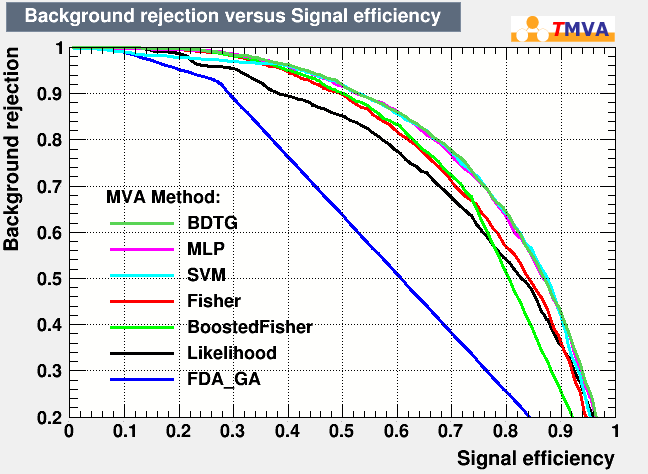}
     \caption{ROC curves showing discriminating power for several TMVA methods when applied to the described binned clustering algorithm variables, for distinguishing voxels in $20\ \text{cm}$ cubes of uranium and lead.} 
     \label{tmva method rocs figure}
  \end{center}    
\end{figure}
Applying the classifier to the training datasets and comparing the resulting AUC for a range of MVA methods (Figure \ref{tmva method rocs figure}) shows that the Gradient-Boosted Decision Tree (BDTG) method is the most suitable, with $\text{AUC}=0.811$. The Multi-Layer Perceptron (MLP) and Support Vector Machine (SVM) methods, which overlap in Figure \ref{tmva method rocs figure}, have AUCs of $0.808$ and $0.804$ respectively. For this reason the BDTG method is used hereafter as the default MVA method.\par
\subsection{Training MVA classifiers}\label{mvatrainsection}
\begin{figure}[htb]  
\begin{center}
      \includegraphics[scale=0.4]{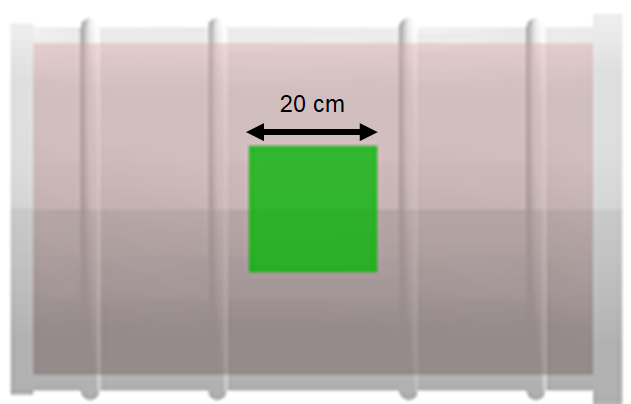}
     \caption{Example simulated geometry used for MVA training: $20\ \text{cm}$ side length uranium cube, in the centre of the waste drum.} 
     \label{U cube training geo figure}
 \end{center}    
\end{figure}

The MVAs were trained on a number of simulated MST muon track data corresponding to a 10 day exposure of four different waste drums: an `empty' drum containing only concrete, and three drums containing $20\ \text{cm}$ side length cubes (see Figure \ref{U cube training geo figure}) of iron, lead and uranium, in the centre of the drum and aligned with its central axis. Only the voxels in the cube (or the equivalent volume for the homogeneous empty drum) were passed to the classifier. The binned cluster algorithm's $n$ parameter (see section \ref{BC algorithm section}) was set to 20. The dataset is split into equally sized `training' and `testing' sets; the MVA is trained on the former then applied to the latter as an overtraining check.
For a binary classifier, one dataset of voxel variables is designated as `signal' and the other `background', whereas a non-binary classifier is passed a single signal dataset and several background datasets. In each case, the classifier attempts to distinguish signal voxels from background(s) voxels, such that when applied to a new voxel it will be classified correctly as often as is possible from the provided variables and the classifier's discriminating power. The non-binary classifiers are trained to distinguish the signal set from all the provided backgrounds (i.e. one-vs-all classification). TMVA calculates an optimum cut value on the classifier response, with a response above the cut being considered `signal-like' and below `background-like'. The optimum cut corresponds to the point at which the signal efficiency is equal to the background rejection. On the ROC curve, this corresponds to the point with the maximum Youden index \cite{youden1950index}, defined as $\text{signal efficiency} + \text{background rejection} -1$; i.e. the length of the vertical line between the ROC curve and the $45^{\circ}$ line connecting the curve's ends.\par
\begin{figure}[htb]  
\begin{center}
      \includegraphics[scale=0.5]{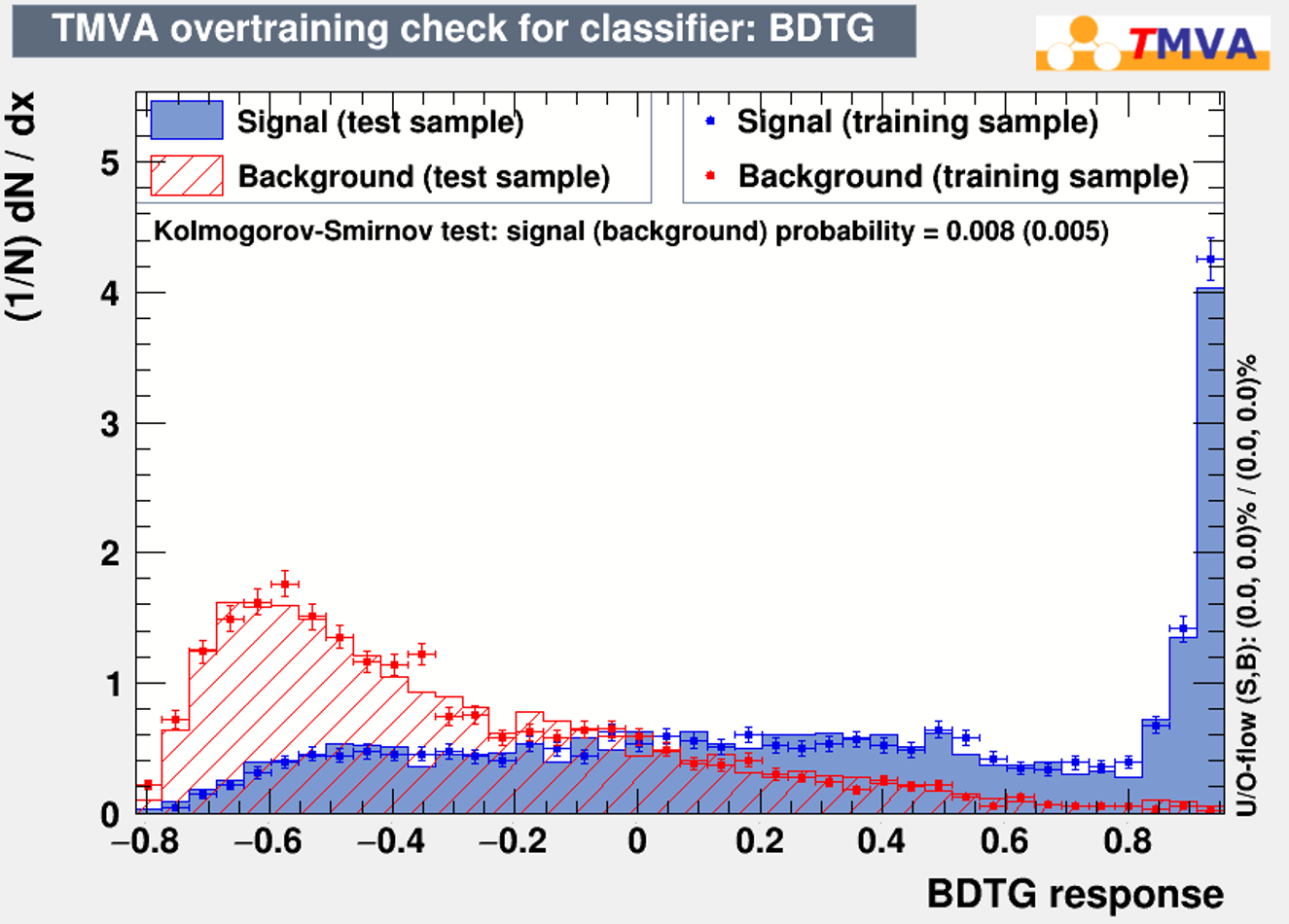} 
     \caption{TMVA overtraining check plot for the uranium-lead binary MVA classifier. The MVA output distributions for the signal and background training sets are overlaid with the output distributions for the test sets for comparison and a Kolmogorov-Smirnov test is performed.}
     \label{overtrainingplot}
 \end{center}    
\end{figure}

To check for overtraining, TMVA's standard check was used: the training signal and background datasets of voxels are both randomly split into two equal groups, with one being used to train the classifier and the other reserved for testing. The trained classifier is then applied to the test set. The classifier output distributions for the training and test sets are then directly compared (see Figure \ref{overtrainingplot}), with a close match between the distributions indicating a low degree of overtraining. A Kolmogorov-Smirnov test is also performed to quantify the similarity of the distributions. In our case, the distributions of the test and training MVA outputs are a close match visually. The Kolmogorov-Smirnov test value is low however, indicating some degree of overtraining has taken place.\par
\subsection{Momentum information} \label{mom_info_section}
\begin{figure}[htb]  
\begin{center}
      \includegraphics[scale=0.3]{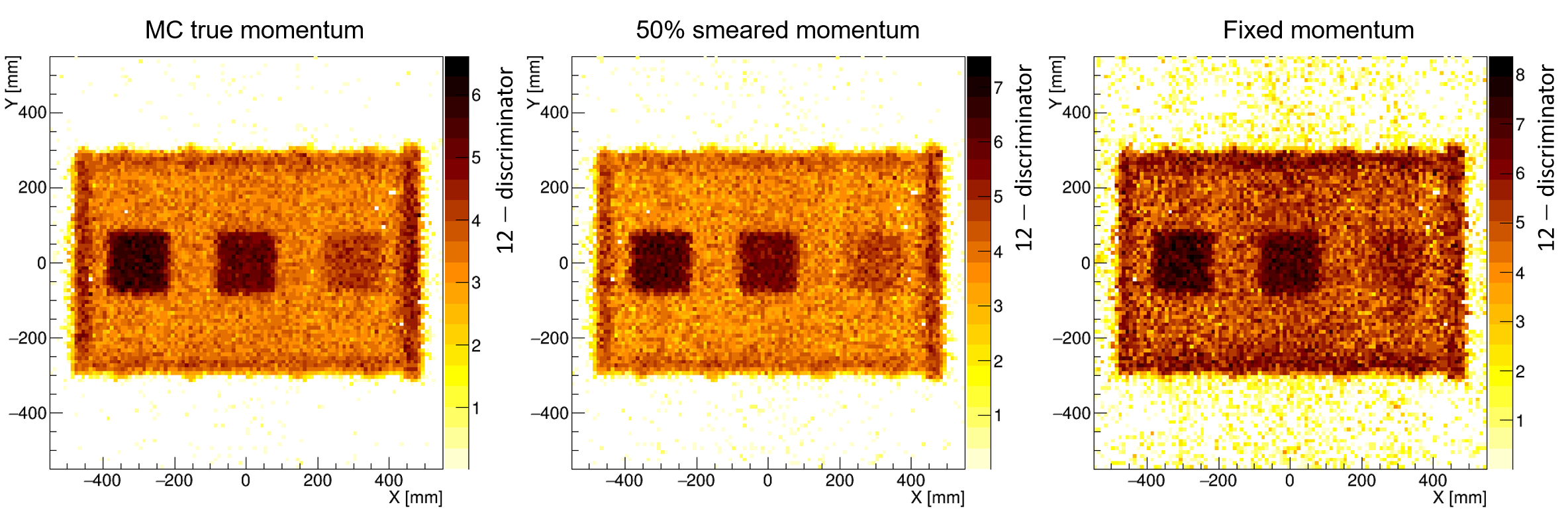}
     \caption{$xy$ slices from binned clustering algorithm output images (with the algorithm's discriminator value subtracted from 12) of a waste drum containing $15\ \text{cm}$ side length cubes of uranium, lead and iron, with three different muon momentum approaches: using the Monte Carlo truth momentum (left), applying a 50\% Gaussian smear to the truth momentum (centre), and removing momentum information entirely by fixing it at a constant value (right). Exposure time = 10 days, $n=5$.} 
     \label{BC image momentum comparison}
 \end{center}    
\end{figure}

To determine the importance of momentum information for material classification, two alternative approaches to the muon momentum were investigated in addition to the 50\% Gaussian smeared truth momentum described in \ref{BC algorithm section}. These were using the Monte Carlo truth momentum itself, with no smearing, and fixing the measured muon momentum at a constant value of $3\ \text{GeV/c}$, i.e. removing momentum information entirely. A comparison of binned clustering algorithm output images of a drum containing $15\ \text{cm}$ cubes of uranium, lead and iron for the different approaches is shown in Figure \ref{BC image momentum comparison}. Using the Monte Carlo truth momentum results in a slightly sharper image with less variation in the concrete background, whereas using fixed momentum significantly reduces the quality of the image with the iron cube in particular difficult to distinguish from the concrete background.\par
\begin{figure}[htb]  
\begin{center}
      \includegraphics[scale=0.4]{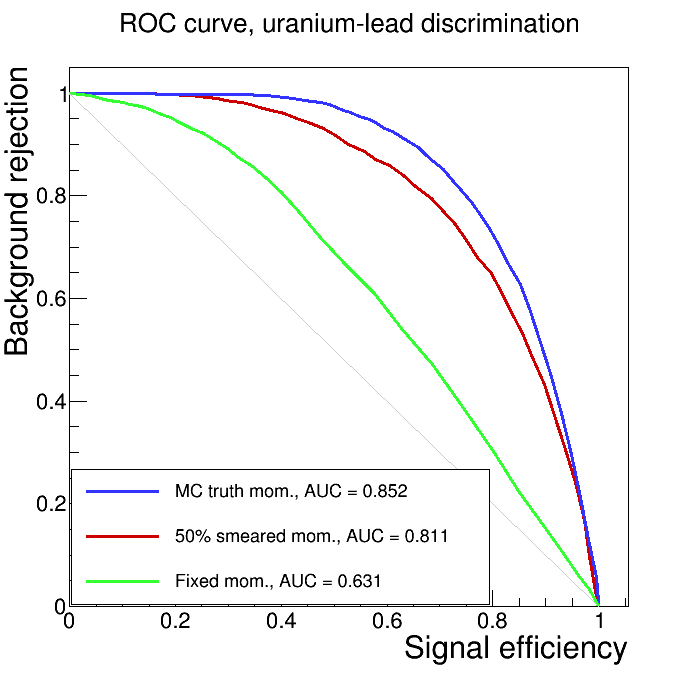}
     \caption{Comparison of ROC curves and their AUCs for the three momentum approaches (Monte Carlo truth momentum, 50\% Gaussian smeared truth momentum, and fixed momentum). The MVA classifier trained to discriminate uranium and lead voxels from samples taken from drums containing $20\ \text{cm}$ cubes, with exposure time 10 days. Smearing the momentum reduces the discriminating power by a small degree, removing momentum information greatly reduces discriminating power.} 
     \label{ROC momentum comparison}
 \end{center}    
\end{figure}

To quantify the effect on material discrimination, binary uranium-lead MVA classifiers trained as described in section \ref{mvatrainsection} but with samples obtained using the three different momentum approaches were used to create ROC curves for each scenario (Figure \ref{ROC momentum comparison}). Comparing the AUC for each case shows that smearing the momentum slightly reduces the discriminating power of the classifier, with $\text{AUC}=0.852$ for the truth momentum and $\text{AUC}=0.811$ for the 50\% smeared momentum. The fixed momentum classifier has significantly worse performance with $\text{AUC}=0.631$. The implication is that momentum information is important for this `local' i.e. voxel-scale approach to material discrimination, but that a smeared momentum approach gives comparable performance to the idealised Monte Carlo truth.\par 
\section{Identifying stored bodies}
\subsection{Removal of concrete background} \label{concrete filtering section}
\begin{figure}[htb]  
\begin{center}
      \includegraphics[scale=0.52]{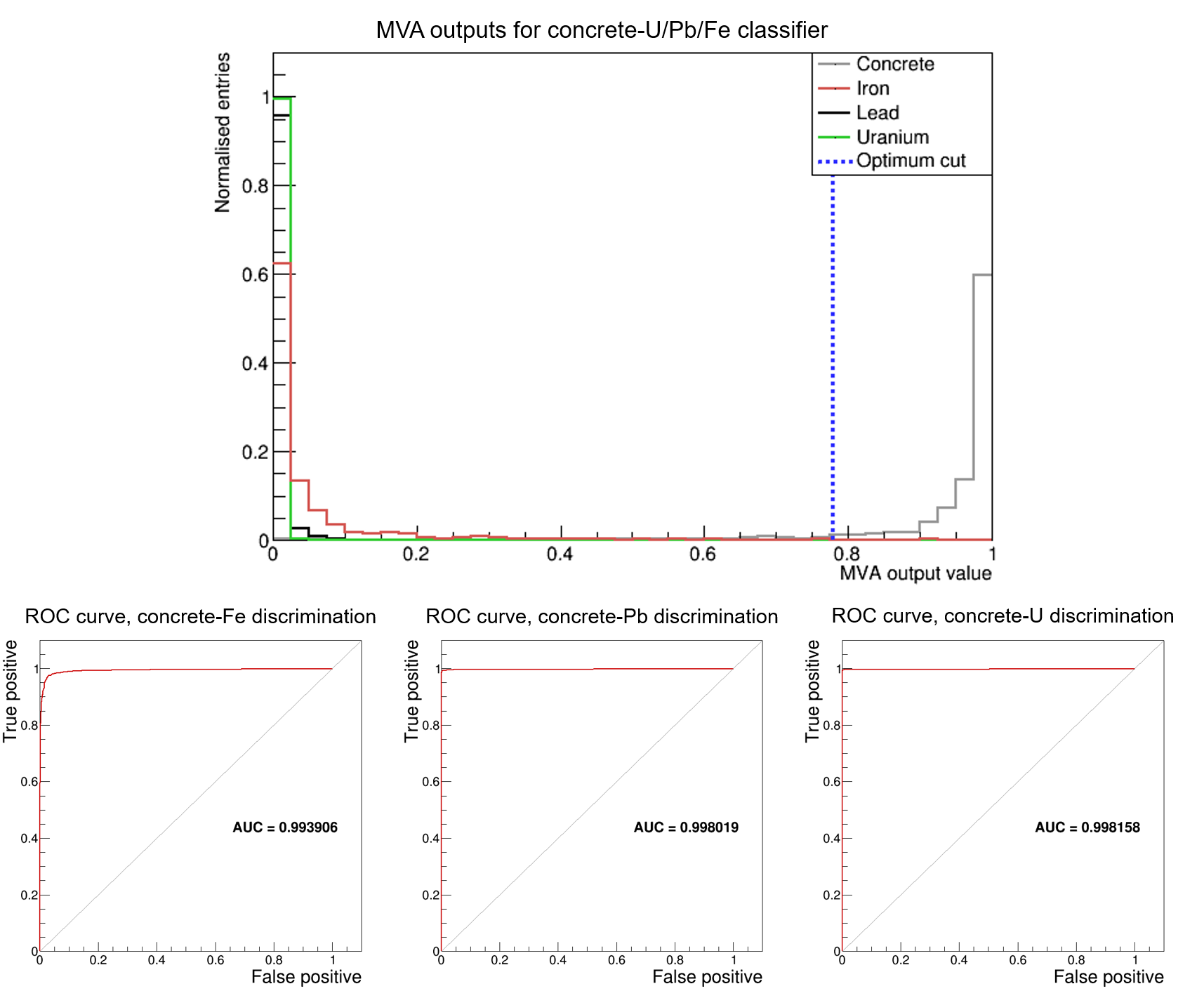}
     \caption{MVA training outputs (top) and ROC curves (bottom) for concrete vs iron/lead/uranium non-binary classifier. The optimum cut (blue) corresponds to the point at which signal efficiency is equal to background rejection.} 
     \label{concrete mva plots figure}
 \end{center}    
\end{figure}

It is necessary to attempt to remove the voxels corresponding to the concrete background and steel shell from the binned clustering algorithm output image. The remaining voxels, corresponding to stored objects, can then be sorted into distinct clusters using the algorithm described in section \ref{clustering section}. The non-binary concrete classifier's training outputs and ROC curves are shown in Figure \ref{concrete mva plots figure}.\par
\begin{figure}[htb]  
\begin{center}
      \includegraphics[scale=0.6]{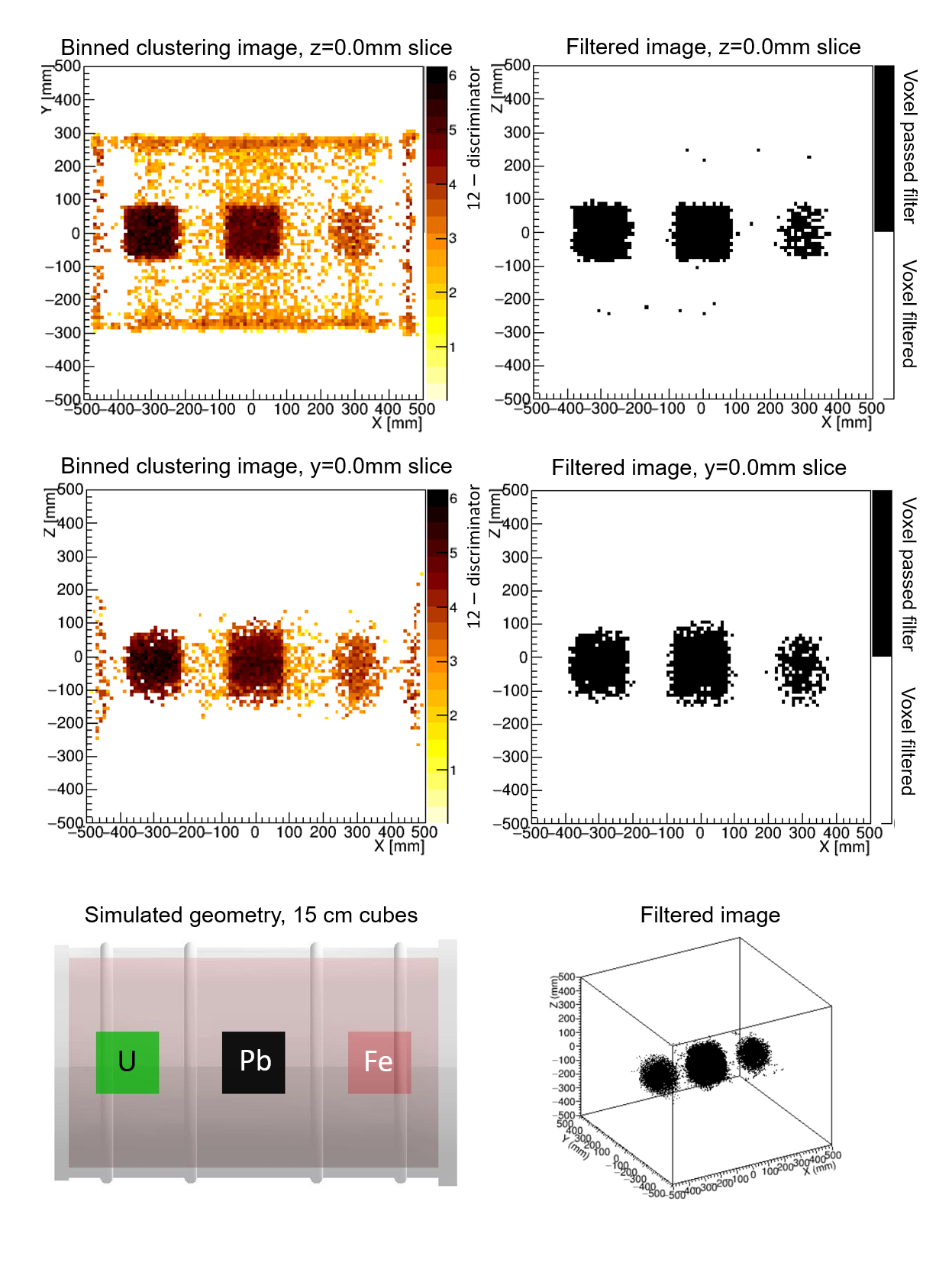}
     \caption{Illustrative example of MVA-filtering algorithm applied to a simulated geometry of a drum containing $15\ \text{cm}$ cubes of uranium, lead and iron. Voxels passing the MVA filtering process described above are coloured black.} 
     \label{MVA_filtered_BC_figure}
 \end{center}    
\end{figure}

As the dimensions of the drum are known, the steel outer shell voxels can be removed trivially thorough a cylindrical spatial cut on the image. Subsequently an MVA classifier trained as described in section \ref{tmva section}, designating the dataset of concrete voxels as `signal' and the other materials as `backgrounds', is applied to the remaining voxels to filter out the concrete voxels. As the classifier is not perfect, some voxels that correspond to concrete in the original simulated geometry remain in the filtered image. The problem is partially mitigated by applying a simple filtering algorithm to remove `isolated' voxels from the image. Each remaining voxel has its 6 nearest neighbour voxels checked; if they are all empty, the voxel is removed from the image. Figure \ref{MVA_filtered_BC_figure} illustrates the result of applying this process to a simulated geometry of three $15\ \text{cm}$ cubes. The removed voxels are coloured white in the images; the remaining voxels are black. To test the performance of the nearest neighbour filtering method, the false positive and false negative rates were calculated for this example. Defining a false positive as voxel that does not correspond to concrete being filtered out, and a false negative as a voxel that does correspond to concrete passing the filter, the false positive rate was $0.014^{+0.008}_{-0.005}$ and the false negative rate was $0.497\pm0.008$. The low false positive rate indicates that very few non-concrete voxels are being incorrectly filtered out. The high false negative rate however indicates that a large number of concrete voxels remain in the final image; this corresponds to the smearing in the $z$ direction of objects in the drum visible in Figure \ref{bc_slice_figure}.\par
\subsection{Clustering} \label{clustering section}
Subsequently these identified and separated `object' voxels need to be grouped into individual clusters, each corresponding to a body stored in the drum. This will allow material information to be calculated by applying MVA classifiers to each identified body. The clustering is achieved through the widely used $k$-means clustering algorithm, which in its simplest form operates as follows:
\begin{itemize}
  \item Choose a value for the number of clusters, $k$.
  \item Pick $k$ randomly selected data points to be the initial cluster centroids.
  \item For each data point, calculate the Euclidean distance (in geometric space) to each of the centroids and assign the point to the cluster with the closest centroid.
  \item Calculate new centroids as the new centres of the clusters.
  \item Repeat until the centroid locations converge.
\end{itemize}
Though this algorithm is fast and easy to implement, it requires the number of clusters $k$ to be known in advance and used as an input. One solution is to run the algorithm multiple times with range of $k$ values as input, and calculate some figure of merit of the clustering output for each. A commonly used figure of merit for clustering algorithms is the Dunn index \cite{dunn1973fuzzy}, defined as the ratio between the minimum inter-cluster distance and the maximum intra-cluster distance. A high Dunn index therefore indicates well-separated and compact clusters. The inter- and intra-cluster distances can be defined to suit the problem; in our case the inter-cluster distance metric is the distance between the closest two data points in the two clusters, and the intra-cluster distance metric is the distance between the two furthest-apart points in a cluster. Defined in this way, the $k$ value that corresponds to the maximum Dunn index will represent the most natural choice for $k$. In most cases this will correspond to the number of bodies stored in the waste drum. In some cases the algorithm can under-estimate $k$ if e.g. two objects are in contact or very close together.\par
\begin{figure}[htb]  
\begin{center}
      \includegraphics[scale=0.75]{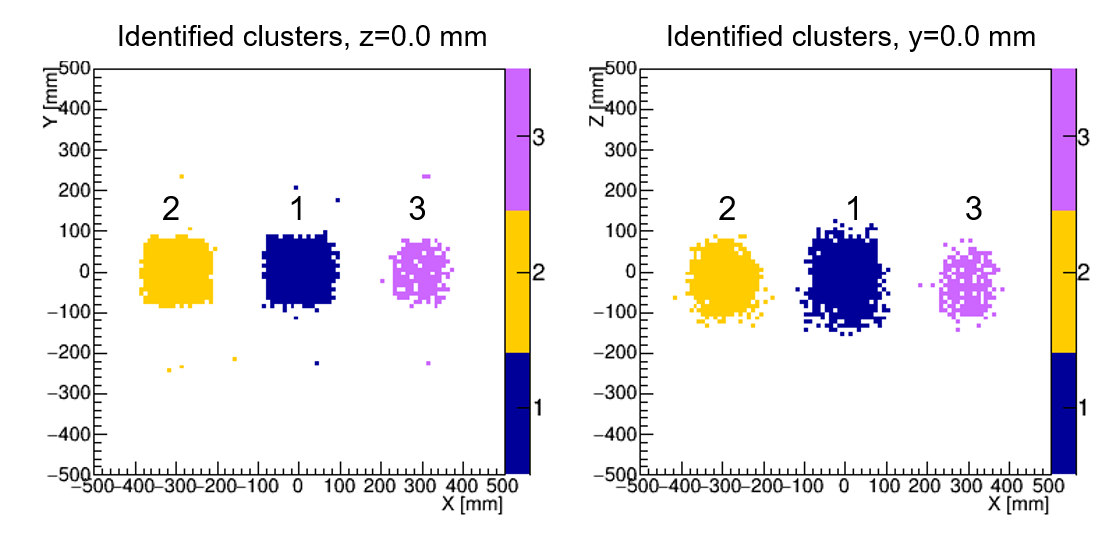}
     \caption{$xy$ (left) and $xz$ (right) slices of the clustering solution for a simulated waste drum containing three $15\ \text{cm}$ cubes of different materials. The voxels separated by the method described in section \ref{concrete filtering section} have been grouped into three clusters using the \textit{k}-means++ clustering algorithm.} 
     \label{cubes15_clustered}
 \end{center}    
\end{figure}

In practise, the simple $k$-means algorithm often produces poor clustering solutions if the randomly chosen initial centroids are too close together. This problem is avoided by choosing the first centroid only from a uniform distribution and the subsequent $k-1$ centroids from a distribution weighted by the squared distances of the data points from the already chosen centroid(s). This form of the algorithm is often referred to as `$k$-means++' \cite{arthur2006k}. Figure \ref{cubes15_clustered} shows the result of applying the $k$-means++ algorithm to a drum containing $15\ \text{cm}$ cubes of iron, lead and uranium.\par
This algorithm occasionally fails when applied to MVA-filtered binned clustering images such as Figure \ref{MVA_filtered_BC_figure}, as the `noise' voxels that do not correspond to a stored object can be treated as a new superfluous cluster. These `fake' clusters are much more sparse than clusters corresponding to stored objects. This allows the problem to be mitigated by defining a cluster density and removing clusters with densities below some cut. We define cluster density as the ratio of the number of voxels in the cluster to the cube of the mean inter-voxel distance. A density cut of $5\times10^{-2}\ \text{voxel}\ \text{cm}^{-3}$ is effective at removing the sparse clusters.\par
A small percentage of voxels that correspond to concrete in the drum will be incorrectly passed by the classifier and included in the filtered image. These will be incorporated into one of the clusters, which could cause an incorrect material decision. These voxels will be outliers in the cluster as the majority of the cluster voxels will be close to the cluster centroid; thus they can be filtered out by placing a cut on the distribution of voxel-centroid distances for each cluster. Choosing the cut so as to remove voxels for which the voxel-centroid distance is greater than one standard deviation from the mean of this distribution is effective at removing outlier voxels.\par
\begin{figure}[htb]  
\begin{center}
      \includegraphics[scale=0.75]{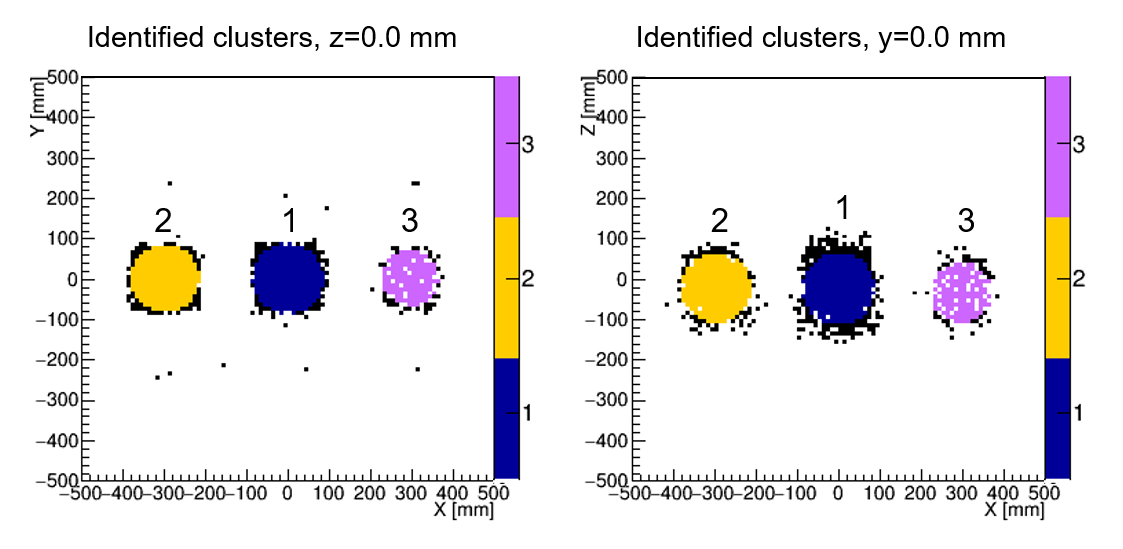}
     \caption{$xy$ (left) and $xz$ (right) slices of the clustering solution of figure \ref{cubes15_clustered} after filtering the outermost voxels from each object. Here black indicates voxels removed from the cluster.} 
     \label{cubes15_outerlayerfiltered}
 \end{center}    
\end{figure}

Finally, a filter is applied to remove approximately the outermost voxel layer (see Figure \ref{cubes15_outerlayerfiltered}) from the surface of each cluster. This is necessary as in general there will be a degree of smearing between a stored body and the concrete background, due to scattering vertices from muons passing close to the object contributing to the algorithm's metric values (see section \ref{BC algorithm section}) and hence affecting the variables that are passed to the MVA classifiers. The filtering is achieved by calculating the mean of the centroid-voxel distances for each cluster, and removing voxels for which the distance is greater than the mean.\par
\section{Results and analysis}
\subsection{Applying MVAs to clustered objects}
\begin{figure}[htb] 
\begin{center}
      \includegraphics[scale=0.6]{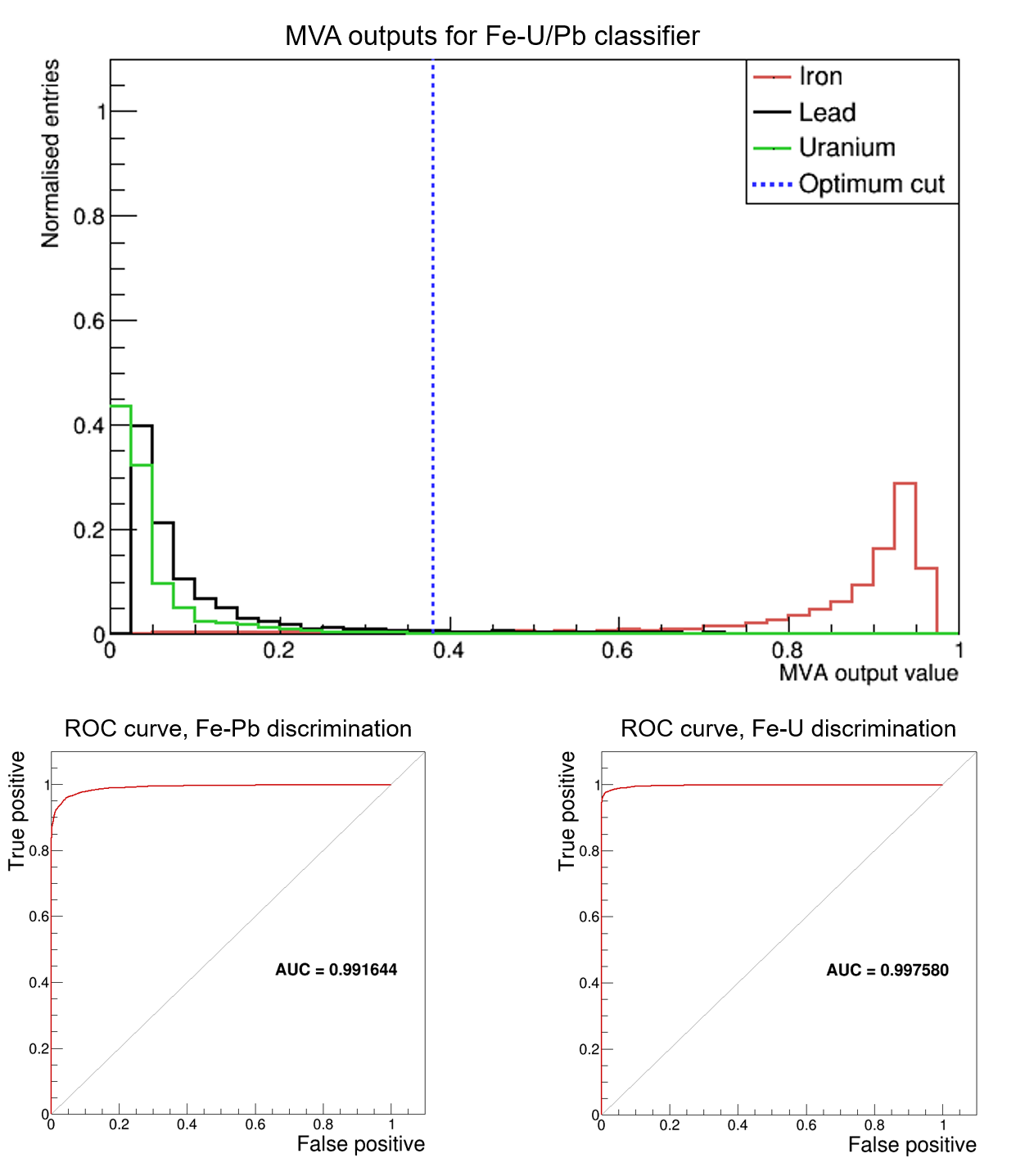}
     \caption{MVA training output and ROC curves for iron/lead/uranium non-binary classifier. The optimum cut corresponds to the point at which signal efficiency is equal to background rejection.} 
      \label{iron_mva_training_figure}
 \end{center}    
\end{figure}

\begin{figure}[htb]  
\begin{center}
      \includegraphics[scale=0.6]{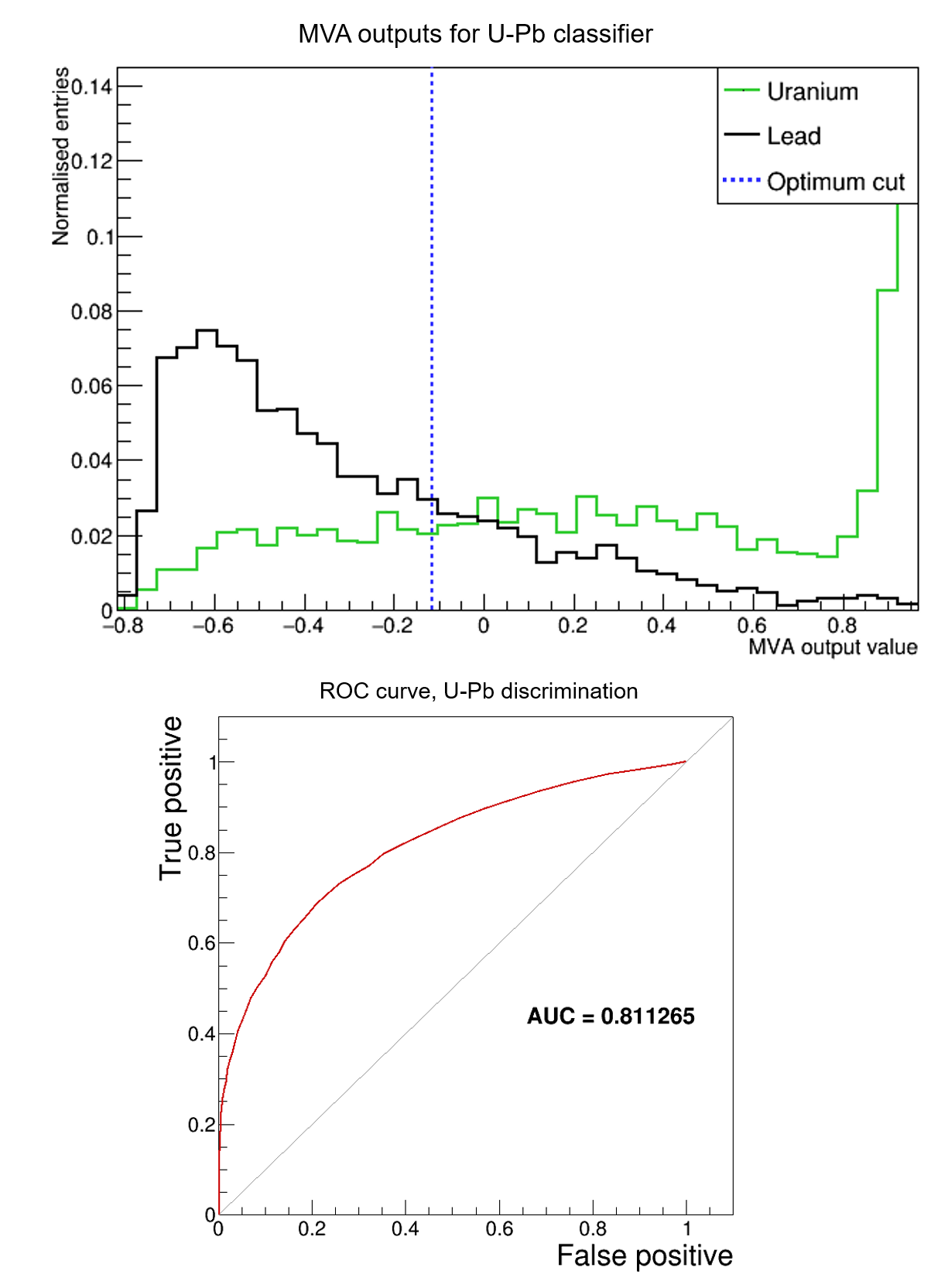}
     \caption{MVA training output and ROC curves for lead/uranium binary classifier. The optimum cut corresponds to the point at which signal efficiency is equal to background rejection.}
     \label{uranium_mva_training_figure}
 \end{center}    
\end{figure}

Further MVA classifiers are now applied to the voxels in each identified cluster to obtain material information for the bodies stored in the drum. Two additional MVA classifiers are trained: a non-binary classifier that separates iron signal from lead and uranium backgrounds (see Figure \ref{iron_mva_training_figure}), and a final binary classifier to discriminate lead and uranium (Figure \ref{uranium_mva_training_figure}). The training ROC AUCs for these classifiers show that the lead and uranium cases are easily distinguished from iron (as the AUC values are close to 1), whereas the lead/uranium classifier does not perform as well, due to the similarity of the materials' $Z$ values.\par
\begin{figure}[htb]  
\begin{center}
      \includegraphics[scale=0.9]{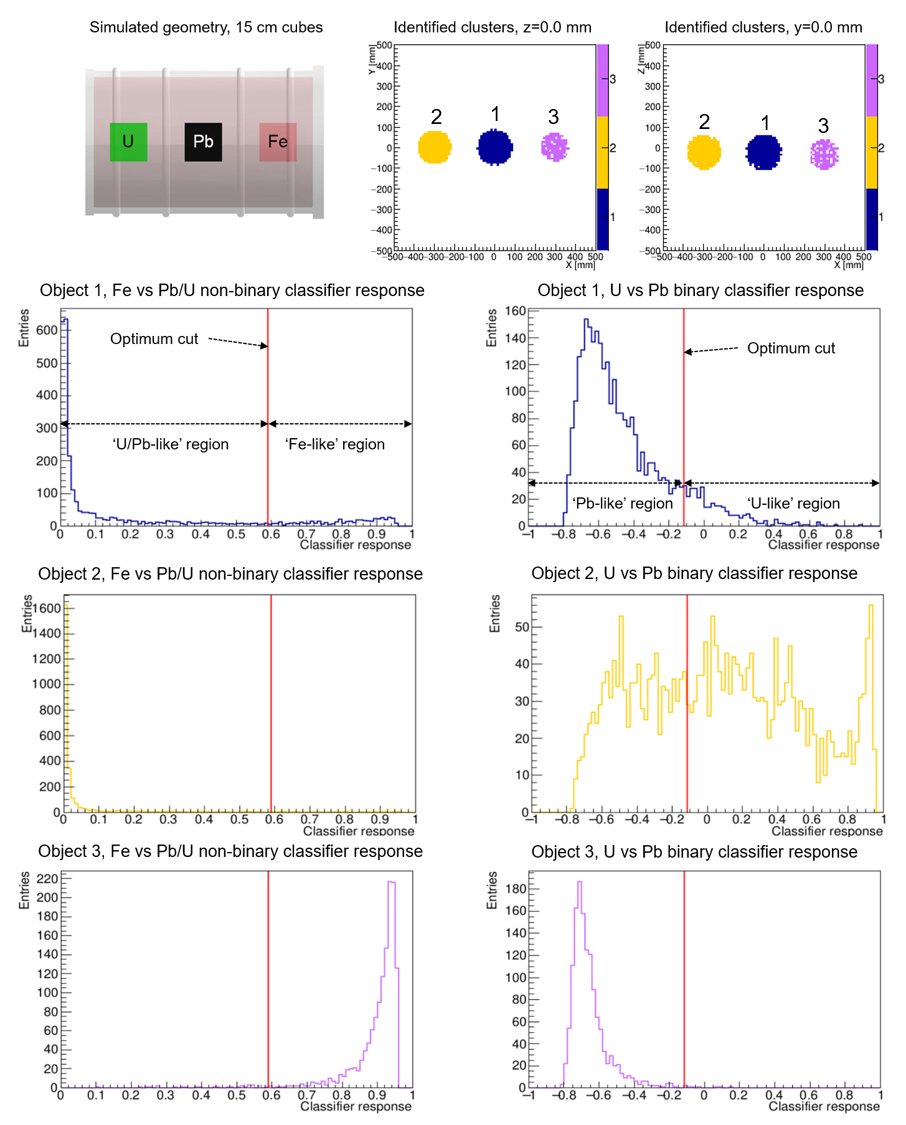}
     \caption{Distributions of responses of MVA classifiers applied to found clusters from a simulated waste drum containing $15\ \text{cm}$ cubes of uranium (object 2), lead (object 1) and iron (object 3). The optimum cuts for the classifiers correspond to the points at which the signal efficiency is equal to background rejection.} 
     \label{cubes15_mva_outputs}
 \end{center}    
\end{figure}

Each MVA classifier will produce a single response value for each voxel it is applied to. If the value falls above the cut (see section \ref{mvatrainsection}), the voxel will be considered signal-like, and if it falls below, background-like. Each identified object is a set of voxels; we apply the classifiers to each voxel to obtain the object's distributions of response values, then calculate the proportions of response values that fall above the cuts (i.e. the proportion of the object's voxels that are signal-like) to arrive at a single value from each classifier for each object. 
Figure \ref{cubes15_mva_outputs} shows the MVA classifier response distributions for the three identified objects in the $15\ \text{cm}$ cube example simulated geometry.\par

\subsection{Obtaining material decisions}
\begin{figure}[htb]  
\begin{center}
      \includegraphics[scale=0.46]{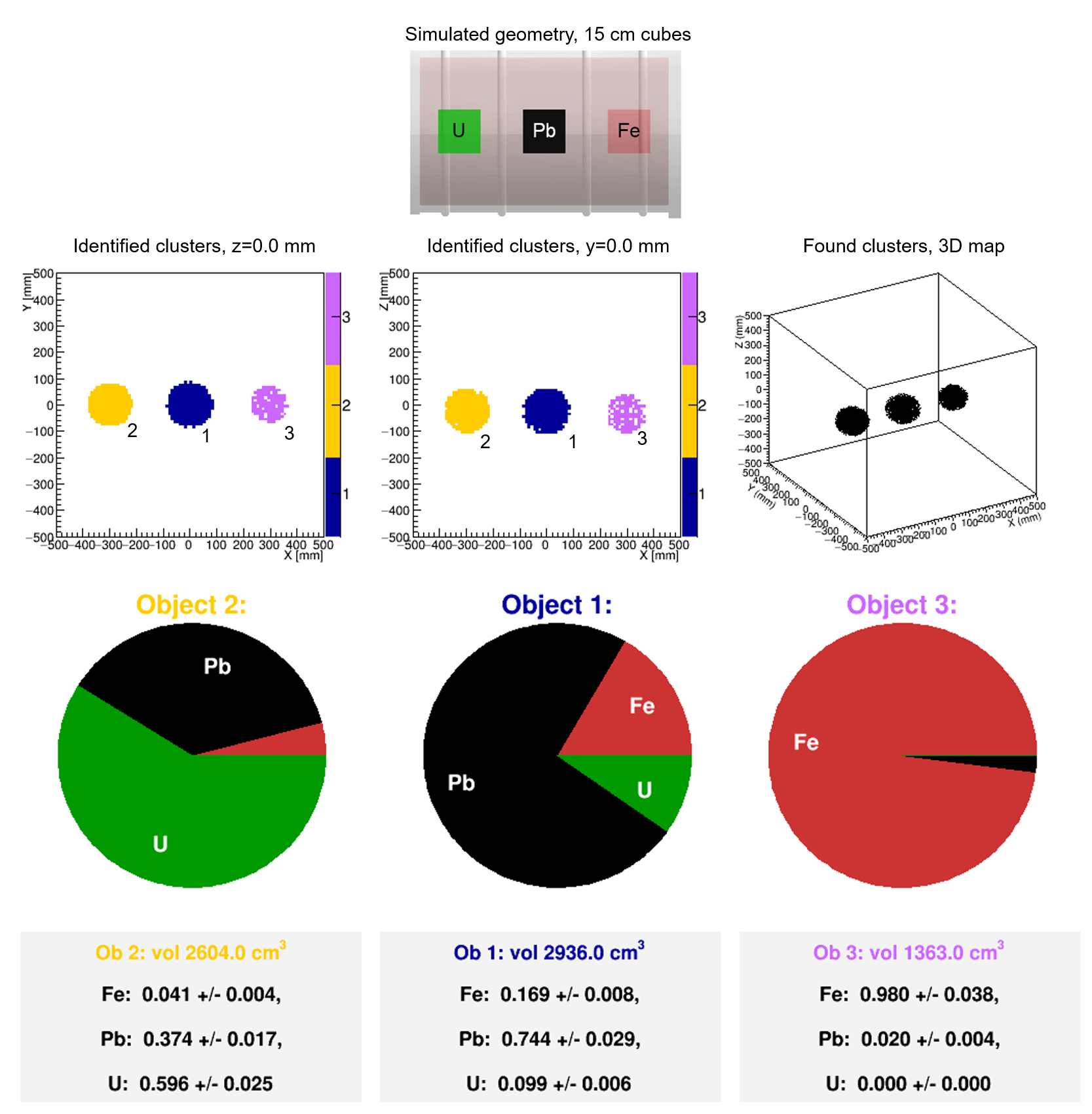}
     \caption{Material scores for simple geometry of three $15\ \text{cm}$ cubes, uranium, lead and iron, aligned with voxel grid.} 
     \label{cubes15_materials}
 \end{center}    
\end{figure}

Applying the integral method described above to these distributions results in uranium, lead and iron `material scores' for each object stored in the drum. The uranium and lead material scores are subsequently multiplied by $1\ -\ \text{iron score}$, i.e. the `not-iron' score. These scores are very effective at distinguishing objects of different materials once the sizes of the objects are taken into account. The material scores are intuitively viewed as a pie chart (see Figure \ref{cubes15_materials}).
For the simulated drum containing three $15\ \text{cm}$ side length cubes of uranium, lead and iron, each object has the MVA-calculated material score that corresponds to the true material as the largest score. The scores for the uranium and lead blocks are also clearly distinguished from each other. However, this simulation is an idealised case due to the large size of the objects and their similarity to the $20\ \text{cm}$ cube training geometries.\par
\begin{figure}[htb]  
\begin{center}
      \includegraphics[scale=0.48]{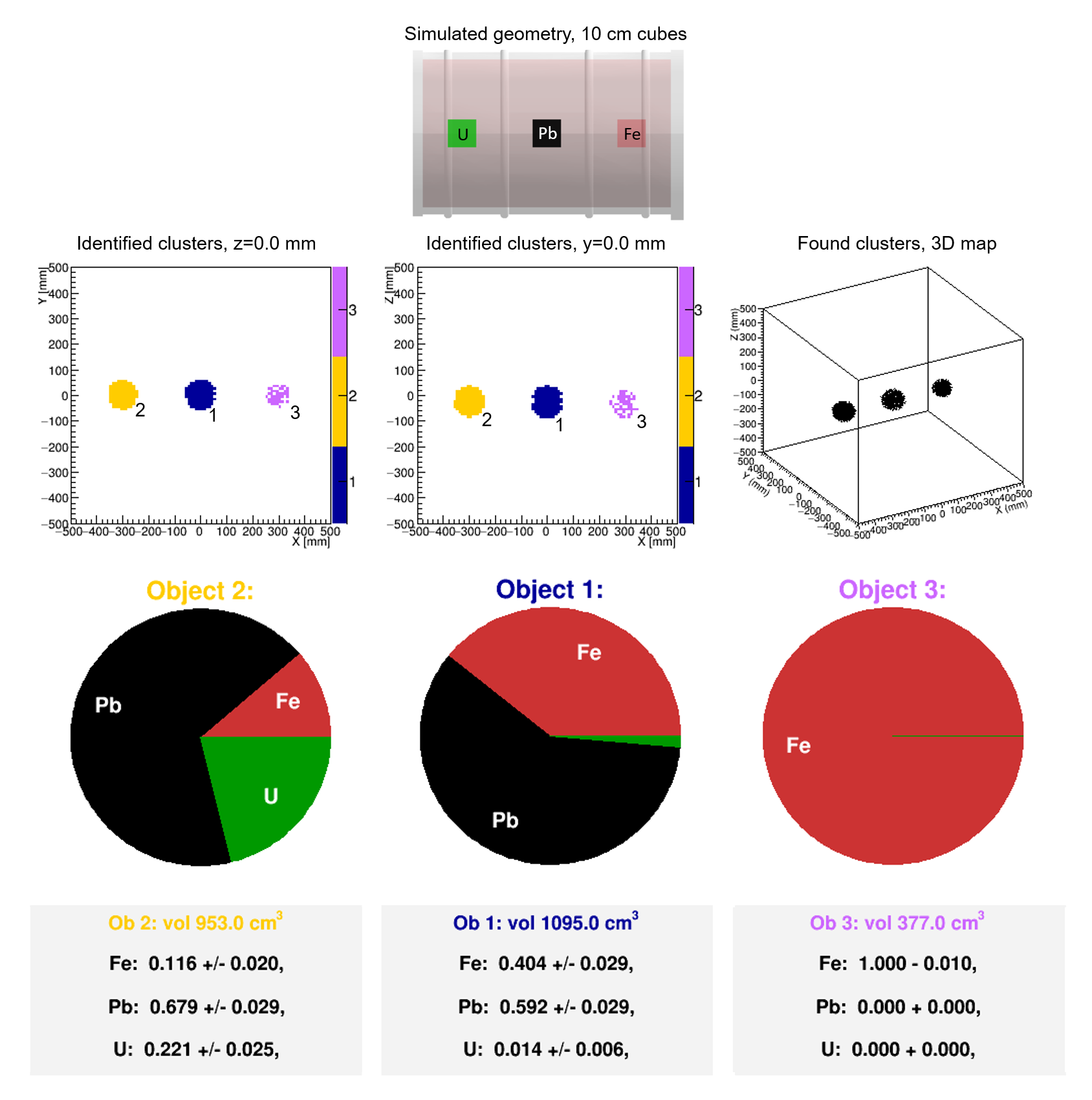}
     \caption{Material estimate results for simple geometry of three $10\ \text{cm}$ cubes, uranium, lead and iron, aligned with voxel grid.} 
     \label{cubes10_materials}
 \end{center}    
\end{figure}

Applying the MVA classifiers to a similar but more challenging geometry of three $10\ \text{cm}$ side length cubes (see Figure \ref{cubes10_materials}), two effects become apparent. Firstly, the classifiers do not perform as well i.e. the score corresponding to the true material is not necessarily the largest. For example, the `uranium' score has reduced from $0.596\ \pm\ 0.025$ for the $15\ \text{cm}$ cube case to $0.221\ \pm\ 0.025$. However, the uranium score for the lead cube has also reduced, by a comparable factor. This effect can be explained by considering the repeated scatterings of muons in a large high-$Z$ object: a larger object will lead to larger detected muon scattering angles, and hence a smaller binned clustering metric value (see \ref{BC algorithm section}). Hence a large lead object can appear more `uranium-like' than a smaller lead object. The implication is that the size of stored objects must be taken into account to reliably determine their material composition.\par
\begin{figure}[htb]  
\begin{center}
      \includegraphics[scale=0.9]{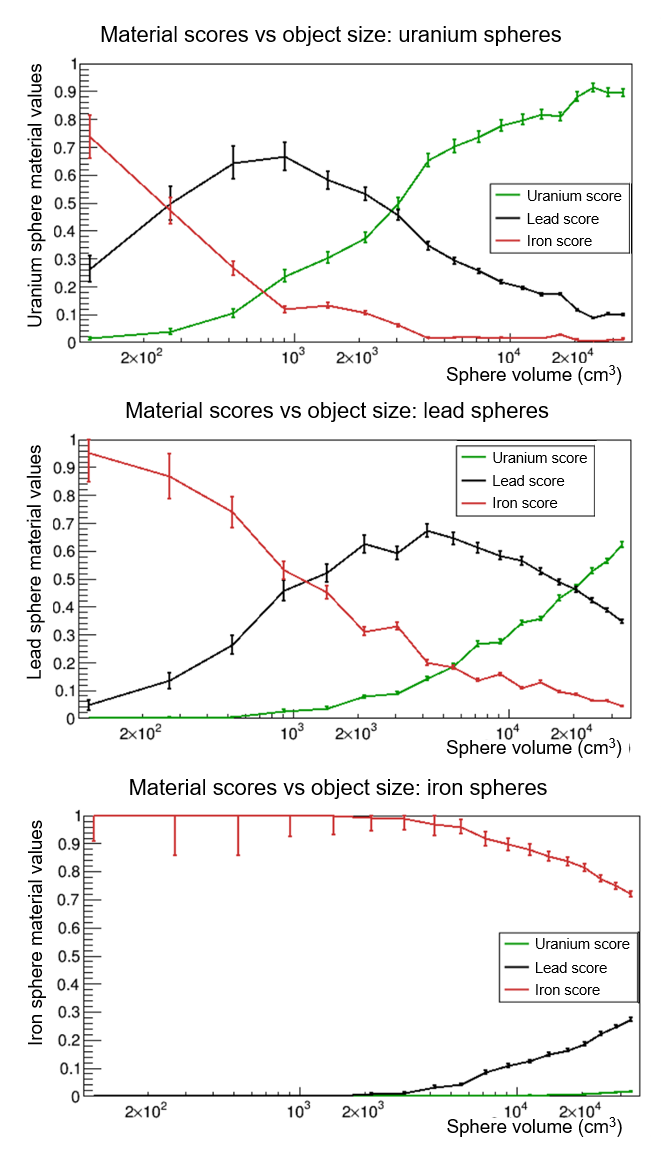}
     \caption{Relationship between the MVA-calculated material scores and the size of the stored object. Each simulated geometry contains a single sphere of increasing radii, composed of uranium (top), lead (middle) or iron (bottom).} 
     \label{sizetest_plot}
 \end{center}    
\end{figure}
To quantify the relations between the object size and the material scores, we applied our system to a series of simulated drums containing spheres of different materials and increasing radii. The results are shown in Figure \ref{sizetest_plot}. It is apparent that whilst there is no simple relation between the material scores and the object volume, objects of different material are clearly distinguished for a wide range of volumes.\par
However, these plots can be used empirically to arrive at a single decision material for each identified stored body in the drum. As the volumes of the clusters (equivalent to the number of constituent voxels) are known, the plots in Figure \ref{sizetest_plot} give the `expected' material scores for a cluster of that size if the object was composed of one of the three materials. Finally, a material decision is arrived at by comparing the object's actual material scores with each set of expected values. The material with the best match, i.e. the minimal 3D Euclidean norm between the actual and expected material scores, is selected as the final material decision.\par
\begin{figure}[htb]  
\begin{center}
      \includegraphics[scale=0.48]{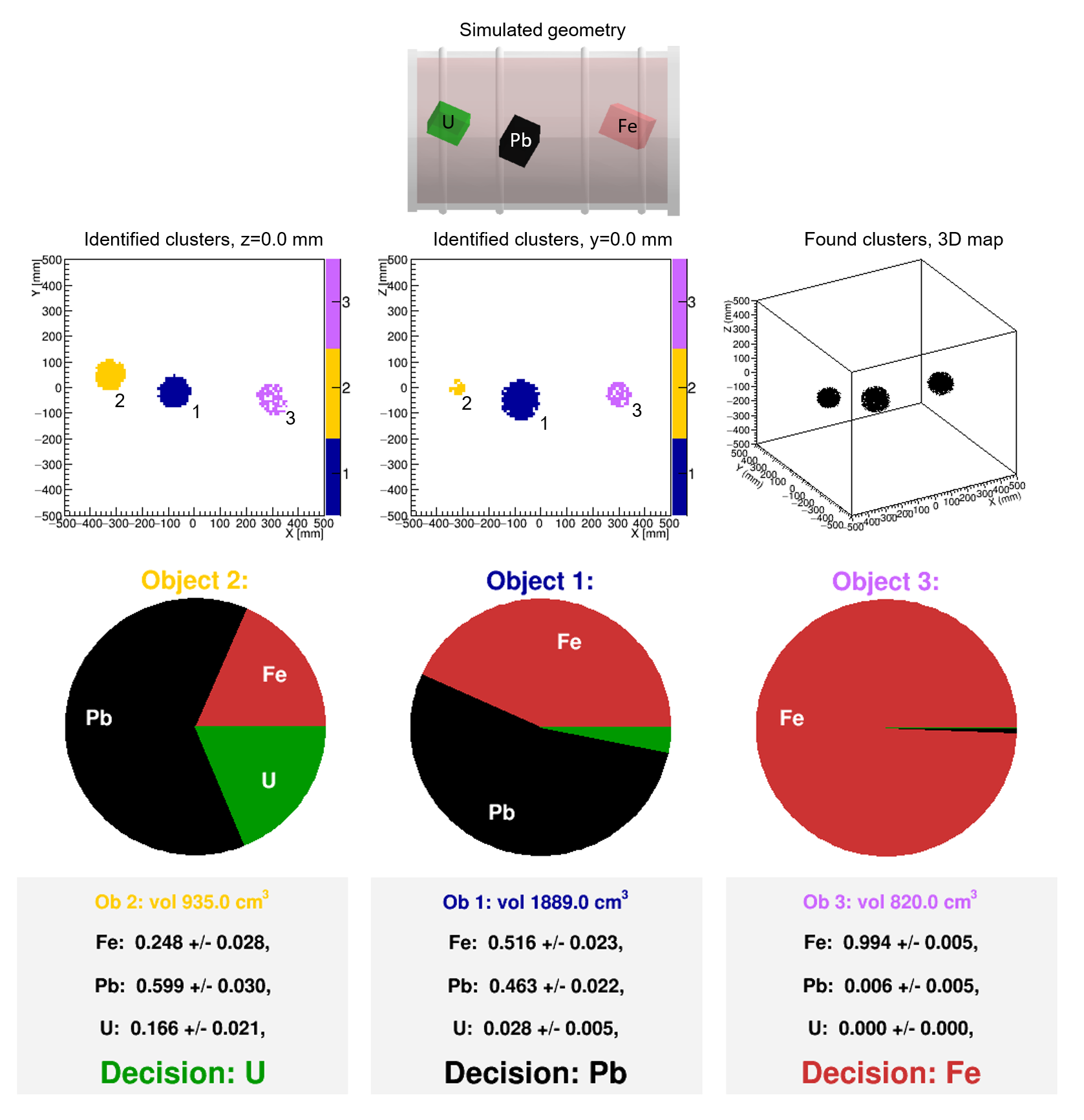}
     \caption{Material estimate results for more complex geometry of three objects, uranium, lead and iron, not aligned with the voxel grid or centred. By calibrating the three material scores against the volume calibration curves, (Figure \ref{sizetest_plot}), the correct material has been assigned in each case.} 
     \label{complex3_materials}
 \end{center}    
\end{figure}

This approach was tested on more complex simulated geometries. Figure \ref{complex3_materials} shows results for a drum similar to the three-cube example of Figure \ref{cubes15_materials}, but with objects of irregular size, location and rotation. In this case the system has accurately identified the correct material for each object. Despite the uranium block's low uranium score compared to the equivalent $15\ \text{cm}$ cube (Figure \ref{cubes15_materials}), the calibration by volume has correctly identified it as uranium.\par
\begin{figure}[htb]  
\begin{center}
      \includegraphics[scale=0.48]{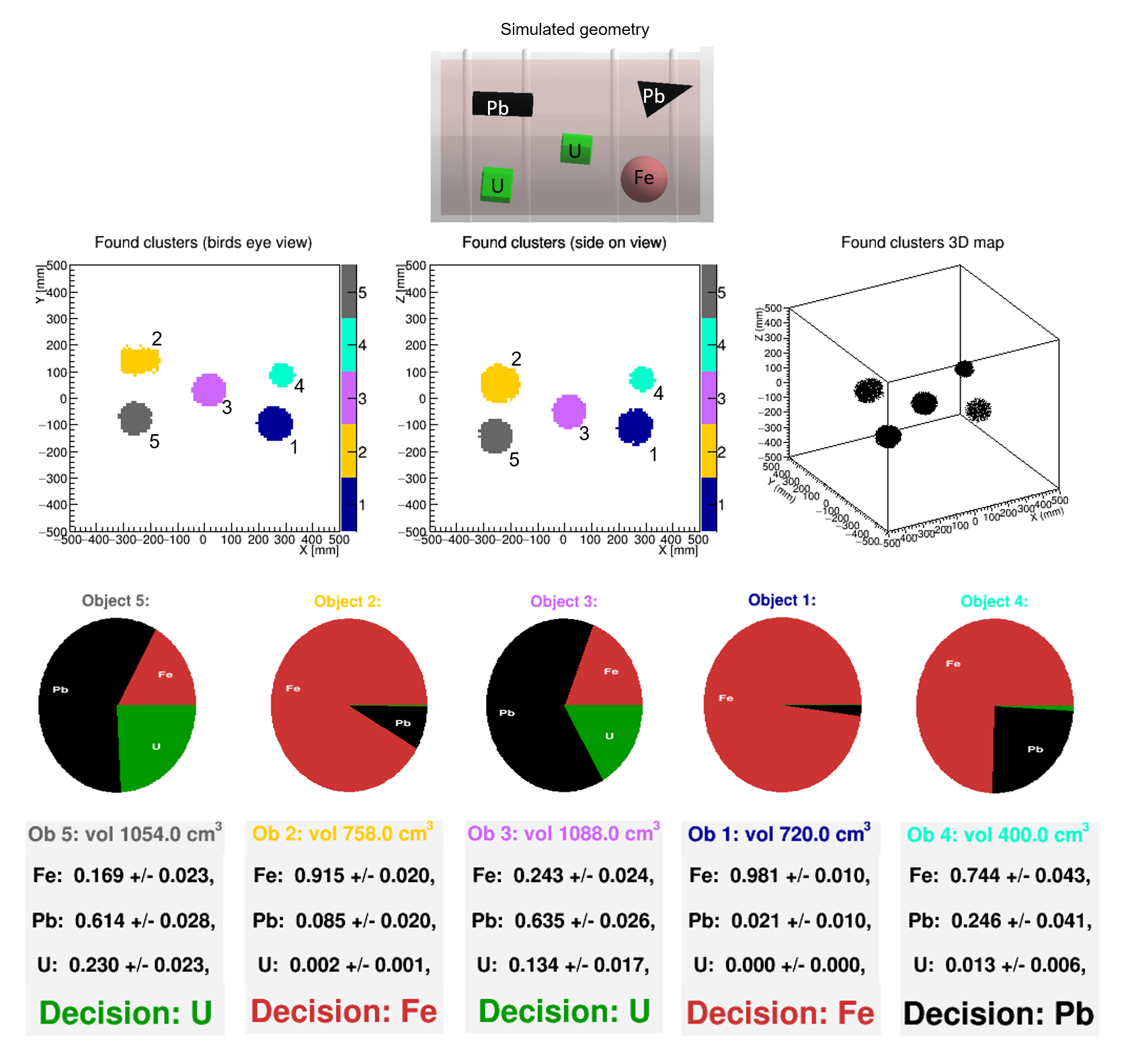}
     \caption{Material estimate results for more complex geometry of five objects of various materials and shapes, dispersed throughout the drum. Note that the 2D cluster plots are viewed as side-on and bird's eye views of the 3D map; this is necessary to view all the clusters as they do not all intersect the central $xy$ and $zx$ planes. Four of the objects have been assigned the correct material; one lead object has been incorrectly classified as iron.} 
     \label{pentuple_materials}
 \end{center}    
\end{figure}

A further example with a larger number of objects is shown in Figure \ref{pentuple_materials}. This drum contains five objects (two uranium, two lead, and one iron) of a wider range of shapes, dispersed more evenly through the drum. However, the system still performs well. The identified clusters are a close match to the true locations of the stored objects. Both uranium objects are correctly assigned, as is the iron sphere and one of the lead objects. One lead object, a tube, has been incorrectly identified as iron. This indicates a limitation of the system when attempting to determine the materials of non-spherical objects.
\subsection{Sensitivity} \label{threat detection section}
To establish the system's sensitivity and false positive rate, we then applied it to a set of randomly generated waste drum simulations. Each simulation contained three spheres of radius $6\ \text{cm}$, randomly dispersed throughout the drum but constrained to not intersect each other. 100 simulations were run in total. 50 simulations contained one uranium, one lead and one iron sphere, and the remaining 50 contained two lead spheres and one iron sphere.  A true positive identification of a uranium object was defined to be an object identified close to the true location of a uranium sphere that was designated as uranium by the system. Conversely a false positive comprised any assignment of a uranium decision to an object in a drum not containing uranium. With these criteria, we found a sensitivity of $0.90^{+0.07}_{-0.12}$, and a false positive rate of $0.12^{+0.12}_{-0.07}$ (95\% Clopper-Pearson confidence intervals).
\section{Conclusions}
We have demonstrated that machine learning techniques are a powerful tool for enhancing the information about a waste drum's contents that can be obtained in a muon scattering tomography experiment. MVA classifiers trained on variables obtained from the distribution of binned clustering algorithm metric values are effective at discriminating materials in waste drums. The concrete matrix can be distinguished from stored objects of mid- and high-$Z$ material, allowing the voxels corresponding to the matrix to be removed, and the remaining object voxels sorted into clusters.\par
Additional material information can be obtained with further MVA classifiers, to discriminate first mid-$Z$ (e.g. iron) from high-$Z$ (lead, uranium) objects, and then between materials with similar $Z$. The effectiveness of the material discrimination is highly dependent on object size. By establishing the empirical relation between object size and the MVA classifiers' material output scores, a final material decision can be made for each identified stored body in the simulated waste drum. This has shown to be accurate for a wide range of object sizes, shapes and drum locations.\par
When tested against a set of simulated drums containing $6\ \text{cm}$ radius spheres of different materials in randomly determined positions, the system performed with a true positive rate of $0.90^{+0.07}_{-0.12}$, and a false positive rate of $0.12^{+0.12}_{-0.07}$, indicating this approach is effective at identifying uranium objects inside waste drums. The main identified vulnerabilities are objects with large differences in $Z$ (e.g. iron and uranium) being very close too each other, and more spatially extended objects being misidentified, although the latter problem could be mitigated by extending the object size-based decision method (see Figure \ref{sizetest_plot}) to account for a wider range of object shapes.\par

\acknowledgments
This project has received funding from the Euratom research and training programme 2014-2018 under grant agreement No 755371.

\newpage

\bibliographystyle{unsrt}
\bibliography{Materials}

\end{document}